\documentclass[aip,jap,reprint,floatfix]{revtex4-2}

\usepackage{graphicx}
\usepackage{dcolumn}
\usepackage{bm}

\usepackage[colorlinks=true, linkcolor=blue, citecolor=red, urlcolor=blue]{hyperref}
\usepackage{paralist}
\usepackage{listings}
\usepackage{pdfsync}
\usepackage{pgfplots}
\usepackage{amsmath}
\usepackage{amssymb}
\usepackage{blindtext}

\usepackage{color}
\definecolor{neworange}{rgb}{1,0.574,0}
\definecolor{darkgreen}{rgb}{0,0.7,0.3}
\usepackage{dcolumn}
\usepackage{fancyhdr}

\newcommand{\blu}{\color{blue} \it }

\begin{document}

\title{Electromagnetic Analogs of Quantum Mechanical Tunnelling}

\author{Jeanne Riga}
\email{jeanne.riga@us.af.mil}
\affiliation{Directed Energy Directorate, Air Force Research Laboratory, Albuquerque, New Mexico   87117,  USA}
\author{Rebecca Seviour}
\affiliation{University of Huddersfield, HD1 3DH, UK}

\date{\today}

\begin{abstract}

In this paper, we introduce the theoretical framework underlying our proposed methodology of verification and validation (V\&V) for quantum mechanical emission models using analogous macroscopic electromagnetic systems.  We derive the correspondence between quantum mechanics and electromagnetism using the transfer matrix approach, and describe the electromagnetic analog that will be used to anchor the atomistic quantum tunneling simulations.  Finally, we illustrate this correspondence by comparing the quantum mechanical and electromagnetic systems for some simple, analytically soluble examples and outline future V\&V work based on the framework presented here.

\end{abstract}


\maketitle
\thispagestyle{fancy}


\section{Introduction}

The physical processes associated with electron emission occur over a wide range of length and time scales concurrently.  While not measured directly by source emission tests in the laboratory, these processes influence the output and performance of macroscopic source emission.  Macroscale material and device-level models developed using classical physics lack the resolution to accurately simulate emission in situations where the electric field is on the order of 5 GeV/m and tunneling through emission barriers on the order of 1 nm in width become significant\cite{Harris2017}.  Fields on this order are typically present at the apex of high ratio emitter such as the fibers on a cathode.  High aspect ratio emitters are necessary to obtain the desired fields at the emission site while keeping fields elsewhere in the device below the breakdown threshold\cite{Harris2018}.  The physical area from which such current is emitted is on the order of tens of square nanometers, and can take the form of protrusions on fibers of micrometer dimension or as unwanted spurious emitters due to surface roughness elsewhere in an High-Power ElectroMagnetic source.

The physics manifested at all length scales impacts the performance of field sources and the overall design of systems in which they will be used.  According to Schottky's conjecture, field enhancements occurring at different length scales combine multiplicatively\cite{Schottky1923, Stern1929, Miller2009, Harris2015, Jensen2016}.  Atomistic effects become even more pronounced if the devices themselves are nanoscale\cite{Kyritsakis2010, Kyritsakis2015, Jensen2017}, which is becoming increasingly common as the push for smaller devices at higher frequencies intensifies.  Even for mesoscale field emitters, nanoscale processes such as erosion\cite{Tang2014}, adsorption of atoms and/or charged inclusions\cite{Binh1992, Zhu2001, Gallmann2017}, and backbombardment/breakdown\cite{Go2010, Edelen2015, Dynako2018} impact the operation of these emitters. Therefore it becomes imperative to have accurate, well-characterized emission physics simulations at the atomistic level in order to accurately predict macroscale electron beam emission.

Previously we have developed a series of trajectory-based approaches to model electrons tunneling through an energy barrier.  A Wigner trajectory formalism was developed and tested by simulating several closed, analytically soluble systems, including an infinite well with infinite and finite barriers separating the halves of the well\cite{Jensen2019, Jensen2019a}. For the infinite half well case, the center barrier was abruptly removed so that the evolution of electron probability density in both halves of the well could be simulated.  For the finite barriers, the probability density in either side of the well was studied for different center barrier heights, Fermi levels, and temperatures.  The methodology was then extended to treat barriers and systems more relevant to field emission, which must be solved numerically because an analytical solution does not exist\cite{Jensen2021}.  The infinite boundaries were replaced by open boundaries, and the transmission and reflection characteristics of a wavepacket incident on Gaussian and parabolic barriers was simulated, along with the delay time associated with barrier interactions, where the solution was found numerically using finite difference methods.  The delay associated with tunneling was studied in more detail in subsequent work \cite{Jensen2021a}.  The McColl-Hartman effect, which characterizes tunneling delay as a function of barrier width, was re-formulated using the Gamow factor, which is a function of barrier shape and incident energy as well as barrier width.  Next, an exact trajectory method based on the Schr\"{o}dinger equation \cite{Jensen2022} was developed to handle systems that are difficult to simulate using the Wigner approach, such as the rectangular and Fowler-Nordheim barriers.  The approaches described here model both above and below barrier emission and delays on an atomistic level, yet the trajectory-based nature of these approaches enables statistical modeling of emission on the mesoscale and generates output that can be used to inform higher-order material and device models.  However, while these models have initially been qualified with analytically soluble examples, it is necessary to anchor non-analytic examples with simulation and experimentation to increase confidence in their use as part of a multiscale modeling tool chain.

In this paper, {we present a theoretical framework illustrating the correspondence between quantum mechanics and electromagnetism, which will be used in subsequent work to verify and validate the aforementioned trajectory simulations}.  While laboratory experiments have been developed to measure tunneling, these experiments are complicated, requiring precise equipment and highly specialized personnel.  Instead, {we propose to use the framework derived herein to verify our atomistic tunneling calculations against simulations of analogous EM systems using well-established EM codes such as HFSS, and to validate the quantum calculations against analogous, EM systems in the RF range using a macroscopic experimental apparatus.  While this initial treatment is not intended to be a verification of our atomistic emission codes, it serves to establish the framework that will be used for V\&V in ongoing and future research}.  The remainder of the paper is organized as follows: in Section II, we present  historical background on the correspondence between QM and EM, in Section III, we show theoretical derivations of the transfer matrices and transmission and reflection coefficients, or S-parameters for both QM and EM as well as deriving the relationship between them, in Section IV we present details of the EM system used to represent the tunneling of a QM particle through a potential barrier, in Section V we show and discuss results for EM and QM treatments of rectangular and delta function barriers, and finally in Section VI we present conclusions and next steps.


 
\section{Background of QM/EM Correspondence}

The concept of using electromagnetics as an analogy for matter waves is deeply embedded in the development of quantum mechanics\cite{Rolf,Shok,Drago}, rooted in a shared  frame work\cite{Ger}, for example, the quantum mechanical wavefunction of a single photon corresponds to the electromagnetic field\cite{Eber}, both of which (quantum mechanical wavefunction and electromagnetic field) can undergo second quantization\cite{Deu}.
 
  It may initially appear to be counter intuitive to draw analogies between photons and electrons, due in part to their differing  properties (rest mass, charge, spin, and EM quantities that are  vectors being scalars in QM\cite{hans}).  However, in cases where electron propagation is well described, {such analogs can be drawn, as in the system in Figure \ref{fig:pb} where wave propagation through a dielectric medium in electromagnetism is analogous to a particle tunneling through a quantum mechanical barrier, $V_b$.} Consider a system described 
  by the single-electron effective mass \cite{Drago}, which is a direct analogy 
 to the Helmholtz equation for the electric field component of an electromagnetic wave in a dielectric medium\cite{lunds}.  We can compare $|E|^2$ of such a system with the probability distribution $|\psi|^2$ in quantum mechanics.  Starting from the time-independent Schrodinger wave equation,  
\begin{equation} \label{eq:eqnrs1}
\nabla^2 \psi +\hat{H} \psi = 0,
\end{equation}
 $\psi$ is the 
 wave function, $\hat{H} = 2m(E-V)/\hbar^2$, $E$ energy and $V$ the potential, and compare with the Helmholtz equation, 
\begin{equation} \label{eq:eqnrs2}
\nabla^2 \bar{\phi} +\hat{T} \bar{\phi} = 0,
\end{equation}
  $\bar{\phi}$ is the 
  electric field and $\hat{T}= (n \omega/c)^2$,  $n$  the refractive index, $c$ the speed of light in vacuum and $\omega$ the angular frequency. The analogy between quantum mechanics and electromagnetics is easily seen when $\hat{H} = \hat {T}$, allowing us to directly express the quantum potential in terms of the equivalent electromagnetic constitutive parameters, 
\begin{equation} \label{eq:eqnrs3}
n = \frac{c_0}{h f} (2m(E_n -V_b))^{1/2},
\end{equation}
 by definition the refractive index ($n$) is given by $n=\sqrt{\epsilon \mu}$, the permittivity ($\epsilon$) and permeability ($\mu$) of the media.  Electromagnetic analogies have been used to study quantum mechanical systems at the macroscopic scale, such as the Dirac point\cite{Hor,Bit}, and to design mesoscopic devices at the macroscopic scale\cite{gay,Drago,glt}. Tunnelling analogys between QM and EM systems have previously been demonstrated in several systems, for example, in multilayer dielectric systems that show the same transmission, reflection and traversal time of electromagnetic waves as the transmission, reflection and traversal time of an electron wavefunction through mesoscopic heterostructures\cite{Drago,gay,glt}. 

The parameters of transmission, reflection and traversal time are three key parameters used when discussing both EM and QM wave propagation across a system. {The analysis in this paper will focus on using the correspondence between quantum mechanics (QM) and electromagnetics (EM) for plane waves incident on simple barriers for analytically soluble systems in order to establish a framework for V\&V of atomistic QM calculations.  We will not consider delays associated with barrier interactions here; this will be addressed in follow-on work where we will use the techniques introduced here to demonstrate V\&V of QM tunneling simulations of quantum wavepackets incident upon more realistic emission barrier shapes.}  

\section{Transfer Matrix Approach}
\begin{figure}
\begin{center}
\includegraphics[width=0.45\textwidth]{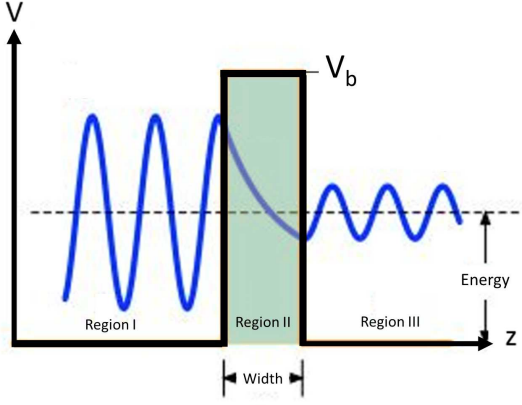}
\caption{\label{fig:pb} Wavefunction for a particle of energy, E, propagating from $-\infty$ and encountering a potential energy barrier of height $V_b$.}
\end{center}
\end{figure}

We begin by reviewing the correspondence between electromagnetic and quantum mechanical tunnelling, comparing the textbook example of a quantum mechanical plane wave incident on a simple rectangular potential barrier to a single frequency (plane wave) of electromagnetic energy propagating through a non-magnetic, isotropic artificial dielectric effective material. The system considered consists of a wave with energy $E$ propagating across three regions, illustrated graphically in Figure \ref{fig:pb}, two semi-infinite regions I \& III, where $E >V_0$, either side of a finite thickness barrier in Region II, where $E<V_b$.  {In this case in Region II, the barrier, the propagation constant of the wavefunction, $\beta$, becomes imaginary and the wave becomes evanescent. Conversely when $E>V_b$ over Region II  the propagation constant $\beta > 0$ is real and the wave propagates sinusoidally over the barrier.}

In this section, we derive the transmission and reflection coefficients via the transfer matrix  for both the EM and QM systems. A more generalized derivation was published previously by Jian et.al. \cite{Jian2018} and the transfer matrix formalism has also been used to model spatially varying potentials \cite{Jensen2018, Ando1998, Mayer2010} as well as references in Ludwick et. al. \cite{Ludwick2021}.  Our example is a specific instance of an isotropic, analytical barrier for the purpose of introducing a concept.  We calculate the transfer matrices for the passage of a single frequency through a dielectric structure for the electromagnetic (EM) example and for the propagation of quantum mechanical plane wave through a rectangular potential barrier for the quantum mechanical (QM) example.  As a further illustration of robustness, a different methodology will be used to calculate the EM and QM transfer matrices.  From the respective transfer matrices, we derive the EM S-parameters and the complex QM reflection (r) and transmission (t) coefficients.  These parameters, along with Eq. \eqref{eq:eqnrs3},  show the correspondence between EM and QM.

\subsection{Electromagnetic Transfer Matrix}
The derivation of the transfer matrix for an EM wave propagating through a dielectric structure has been shown previously \cite{Berreman1972, Rogers2011, Figotin2001, Figotin2003}.  We begin with the differential form of Maxwell's equations in the Gaussian formulation in the absence of charge and sources:
\begin{equation} \label{eq:Maxwell}
\begin{array}{c}
\nabla \cdot \bar{E} = 0, \>\> \nabla \cdot \bar{B} = 0\\ \\
\nabla \times \bar{E} = -\frac{1}{c}\frac{\partial\bar{B}}{\partial t}\\ \\
\nabla \times \bar{H} = \frac{1}{c}\frac{\partial\bar{D}}{\partial t}
\end{array}
\end{equation}
where $\bar{B}=\hat{\mu}\bar{H}$ and \textcolor{red}{$\bar{D}=\hat{\epsilon}\bar{E}$}, the magnetic and electric fields within a non-vacuum medium.  The field vectors in this instance are plane waves:
\begin{equation} \label{eq:Efield}
\bar{E}(z,t)=\bar{E}_0\,e^{i(kz-\omega t)}+\bar{E}_0\,e^{-i(kz+\omega t)}
\end{equation}
\begin{equation} \label{eq:Hfield}
\bar{H}(z,t)=\bar{H}_0\,e^{i(kz-\omega t)}+\bar{H}_0\,e^{-i(kz+\omega t)}
\end{equation}
where the wavenumber, $k=\frac{\omega}{c}n$, $n$ is the index of refraction: $n=\sqrt{\epsilon \mu}$.  For the purpose of this simple illustration, we assume a lossless medium so the imaginary components of $\epsilon$, and $\mu$ are zero and that the plane wave is propagating in the positive z-direction such that it is normally incident on an isotropic dielectric medium.  We also assume the electric and magnetic field components align with the x and y axes, respectively.  The field vectors can then be rewritten as:
\begin{equation} \label{eq:Evector}
\bar{E}(z,t)= \begin{bmatrix}E_x(z,t)\\E_y(z,t)\end{bmatrix} = \begin{bmatrix}E_x e^{ikz} + E_x e^{-ikz}\\E_y e^{ikz} + E_y e^{-ikz}\end{bmatrix}\,e^{-i\omega t}
\end{equation}
\begin{equation} \label{eq:Hvector}
\bar{H}(z,t)=\begin{bmatrix}H_x(z,t)\\H_y(z,t)\end{bmatrix} =\begin{bmatrix}H_x e^{ikz} + H_x e^{-ikz}\\H_y e^{ikz} +H_y e^{-ikz}\end{bmatrix}\,e^{-i\omega t}
\end{equation}
with the cumulative field vector represented as:
\begin{equation} \label{eq:psivector}
\psi(z,t)=\begin{bmatrix}E_x(z,t)\\E_y(z,t)\\H_x(z,t)\\H_y(z,t)\end{bmatrix}
\end{equation}
When either of the non-zero Maxwell's equations operate on the electric and mangnetic field vectors, 2 field polarizations result, each with a forward and reverse component:
\begin{equation} \label{eq:pol1}
\psi_1(z,t)=\left(\begin{bmatrix}Z\\0\\0\\1\end{bmatrix}\cdot e^{ikz}+\begin{bmatrix}-Z\\0\\0\\1\end{bmatrix}\cdot e^{-ikz}\right)\cdot e^{-i\omega t}
\end{equation}
\begin{equation} \label{eq:pol2}
\psi_2(z,t)=\left(\begin{bmatrix}0\\-Z\\1\\0\end{bmatrix}\cdot e^{ikz}+\begin{bmatrix}0\\Z\\1\\0\end{bmatrix}\cdot e^{-ikz}\right)\cdot e^{-i\omega t}
\end{equation}
where Z is the normalized impedance of the propagation medium.  In analogy with the time-independent Schr\"{o}dinger equation, the spatial parts of these forward and reverse eigenvectors are then used to construct the transfer matrix in the following way for propagation through a dielectric slab of thickness $L$ \cite{Figotin2001}:
\begin{equation} \label{eq:TW}
\hat{T}(z,0)=\hat{W}(z)\,\hat{W}^{-1}(0)
\end{equation}
where the columns of the 4x4 $\hat{W}(z)$ matrix are simply the forward and reverse field vectors within the dielectric medium for each of the polarizations above:
\begin{equation} \label{eq:Wmatrix}
\hat{W}(z)=\begin{pmatrix}Z_d\,e^{iqz} & -Z_d\,e^{-iqz} & 0 & 0 \\ 0 & 0 & -Z_d\,e^{iqz} & Z_d\,e^{-iqz} \\ 0 & 0 & e^{iqz} & e^{-iqz} \\ e^{iqz} & e^{-iqz} & 0 & 0 \end{pmatrix}
\end{equation}
where the wavevector within the dielectric medium, $q=\frac{\omega}{c}n_d$, $n_d$ is the normalized refractive index of the dielectric medium, and $Z_d$ is the normalized impedance of the dielectric medium.  Taking the inverse of this matrix at z=0, we get:
\begin{equation} \label{eq:W0}
\hat{W}^{-1}(0)=\begin{pmatrix} \frac{1}{2 Z_d} & 0 & 0 & \frac{1}{2} \\ \frac{-1}{2 Z_d} & 0 & 0 & \frac{1}{2} \\ 0 & \frac{-1}{2 Z_d} & \frac{1}{2} & 0 \\ 0 & \frac{1}{2 Z_d} & \frac{1}{2} & 0 \end{pmatrix}
\end{equation}
Multiplying the matrices in Eqs. \eqref{eq:Wmatrix} and \eqref{eq:W0} together, we obtain the transfer matrix for a dielectric slab of thickness L:
\begin{equation} \label{eq:TmatrixEM}
\begin{split}
&\hat{T}(L,0) = \\ &\begin{pmatrix} \cos(qL) & 0 & 0 & iZ_d\sin(qL) \\ 0 & \cos(qL) & -iZ_d\sin(qL) & 0 \\ 0 & \dfrac{-i}{Z_d} \sin(qL) & \cos(qL) & 0 \\ \dfrac{i}{Z_d} \sin{(qL)} & 0 & 0 & \cos(qL) \end{pmatrix}
\end{split}
\end{equation}
We note  that this 4x4 transfer matrix is the general transfer matrix expression that applies to any medium, including anisotropic media where orthogonal field polarizations experience different environment and propagation characteristics in the medium.  For the isotropic material considered here, however, both polarizations are equivalent, so one of them can be chosen arbitrarily.  Since the non-zero terms of a given polarization vector interact with only 4 of the non-zero elements of the transfer matrix, the dimensionality of both the vector and matrix can be reduced as shown for Polarization 2:
\begin{equation} \label{eq:4to2}
\begin{split}
&\psi_2(0,0) = \begin{bmatrix}0\\-Z_d\\1\\0\end{bmatrix} \Rightarrow \begin{bmatrix}-Z_d\\1\end{bmatrix} \\
&\hat{T}(L,0) = \\ &\begin{bmatrix} \cos(qL) & 0 & 0 & iZ_d\sin(qL) \\ 0 & \cos(qL) & -iZ_d\sin(qL) & 0 \\ 0 & \frac{-i}{Z_d} \sin(qL) & \cos(qL) & 0 \\ \frac{i}{Z_d} \sin(qL) & 0 & 0 & \cos(qL) \end{bmatrix}\\ 
&\Rightarrow \begin{bmatrix}\cos(qL)&-iZ_d\sin(qL)\\-\frac{i}{Z_d}\sin(qL)&\cos(qL)\end{bmatrix}
\end{split}
\end{equation}
A convenient way of representing the function of the electromagnetic transfer matrix is \cite{Figotin2001, Figotin2003}:
\begin{equation} \label{eq:Tprop}
    \psi(z)=\hat{T}(z,z_0)\psi(z_0)
\end{equation}
where the thickness of dielectric medium is $z-z_0$.  In our example of a dielectric slab of thickness L,
\begin{equation} \label{eq:irt}
    \psi(z)=t\psi_t(L)\,\,\,\,\,\,\psi(z_0)=\psi_i(0)+r\psi_r(0)
\end{equation}
where $r$ and $t$ are the transmission and reflection coefficients that we will calculate, and the incident, reflected, and transmitted vectors in equation \eqref{eq:irt} are those propagating through vacuum on either side of the dielectric slab.  The field vector propagates through free space on either side of the slab, and as there is only vacuum on the right hand side of the slab, the transmitted component will have only forward propagating components.  Combining Eqs. \eqref{eq:Tprop} and \eqref{eq:irt} and writing them out explicitly for polarization 2, we get:
\begin{equation} \label{eq:Z0form}
\begin{bmatrix}-tZ_0 e^{ikz} \\ te^{ikz}\end{bmatrix} = \hat{T}(L,0) \begin{bmatrix} -Z_0e^{ikz}+rZ_0e^{-ikz} \\ e^{ikz}+re^{-ikz} \end{bmatrix}
\end{equation}
where $Z_d$ is the normalized impedance of the dielectric medium and $Z_0$ is the normalized impedance of free space.  The left boundary of the dielectric is at $z=0$, while the right hand boundary of the dielectric is at $z=L$.  Substituting the numerical values for impedance and position on either side of the barrier, the equation is rewritten as:
\begin{equation} \label{eq:MatrixAlgebra}
\begin{bmatrix}-te^{ikz} \\ te^{ikz}\end{bmatrix} = 
\hat{T}(L,0) \begin{bmatrix} -1+r\\ 1+r \end{bmatrix}
\end{equation}
From this equation, we derive expressions for the reflection and transmission parameters, $r$ and $t$, which are the same as the EM S-parameters:
\begin{align} 
\label{eq:S11pos}
r=S_{11}&=\frac{i(Z_d-\frac{1}{Z_d})\sin(qL)}{2\cos(qL)-i(Z_d+\frac{1}{Z_d})\sin(qL)} \\
\label{eq:S21pos}
t=S_{21}&= \frac{2e^{-ikL}}{2\cos(qL)-i(Z_d+\frac{1}{Z_d})\sin(qL)}
\end{align}
Since this material is isotropic, calculation of the transmission and reflection coefficients using the other polarization will yield identical results.  Eqs. \eqref{eq:S11pos} and \eqref{eq:S21pos} are what would result for a dielectric material with a real, positive index of refraction, which corresponds to above barrier propagation in quantum mechanics.  The analogy to below barrier tunneling occurs for a material with a negative refractive index, where the oscillating field vector becomes evanescent within the material. This occurs, for example, in materials where $\epsilon <0$ and $\mu>0$.  The wavenumber within the material becomes imaginary, $q\rightarrow i\rho$, and the S-parameter expressions become:
\begin{align} 
\label{eq:S11neg}
S_{11} &= \frac{-(Z_d-\frac{1}{Z_d})\sinh(\rho L)}{2\cosh(\rho L)+(Z_d+\frac{1}{Z_d})\sinh(\rho L)} \\
\label{eq:S21neg}
S_{21} &= \frac{2e^{-ikL}}{2\cosh(\rho L)+(Z_d+\frac{1}{Z_d})\sinh(\rho L)}
\end{align}

\subsection{Quantum Mechanical Transfer Matrix}
We  now present the derivation of the QM transfer matrix for above barrier propagation.  The wavefunction is propagating in free space in the positive z-direction from $-\infty$, and it encounters a barrier of height $V_b$ at $z=0$ having thickness $L$.  We begin with the spatial components of the wavefunction \cite{Cohen-Tannoudji1977},\cite{Cahay2017a} in each of the 3 regions shown in Figure \ref{fig:pb} :
\begin{align} 
\label{eq:psi1}
\psi_1(z)&=A_1e^{ik_1z}+A'_1e^{-ik_1z} \\
\label{eq:psi2}
\psi_2(z)&=A_2e^{ik_2z}+A'_2e^{-ik_2z} \\
\label{eq:psi3}
\psi_3(z)&=A_3e^{ik_3z}+A'_3e^{-ik_3z}
\end{align}
where regions 1 and 3 are free space, and region 2 is the barrier region.  The wavenumbers in each of these regions are:
\begin{equation} \label{eq:kfree}
k_1=k_3=k=\sqrt{\frac{2mE}{\hbar^2}}
\end{equation}
\begin{equation} \label{eq:qbarrier}
k_2=q=\sqrt{\frac{2m(E-V_b)}{\hbar^2}}
\end{equation}
where E is the energy of the plane wave and $V_b$ is height of the potential energy barrier.  The derivation will be carried out for the case of above barrier propagation, $E>V_b$.  The free space wavenumber is k, and the wavenumber in the barrier region is q.

According to the boundary conditions for quantum mechanics, both the wavefunction and its derivative must be continuous at the interfaces, so at $z=0$, $\psi_1(z)=\psi_2(z)$ and $\psi'_1(z)=\psi'_2(z)$ and at $z=L$, $\psi_2(z)=\psi_3(z)$ and $\psi'_2(z)=\psi'_3(z)$.  Putting the boundary conditions for the first interface into matrix form:
\begin{equation}
\begin{bmatrix}e^{ikz}&e^{-ikz}\\ik\,e^{ikz}&-ik\,e^{-ikz}\end{bmatrix}\begin{bmatrix}A_1\\A'_1\end{bmatrix}=\begin{bmatrix}e^{iqz}&e^{-iqz}\\iq\,e^{iqz}&-iq\,e^{-iqz}\end{bmatrix}\begin{bmatrix}A_2\\A'_2\end{bmatrix}
\end{equation}
and then setting $z=0$, we get:
\begin{equation} \label{eq:boundaryz0}
\begin{bmatrix}1&1\\ik_1&-ik_1\end{bmatrix}\begin{bmatrix}A_1\\A'_1\end{bmatrix}=\begin{bmatrix}1&1\\iq&-iq\end{bmatrix}\begin{bmatrix}A_2\\A'_2\end{bmatrix}
\end{equation}
The boundary condition for the second interface is:
\begin{equation}
\begin{bmatrix}e^{iqz}&e^{-iqz}\\iq\,e^{iqz}&-iq\,e^{-iqz}\end{bmatrix}\begin{bmatrix}A_2\\A'_2\end{bmatrix}=\begin{bmatrix}e^{ikz}&e^{-ikz}\\ik\,e^{ikz}&-ik\,e^{-ikz}\end{bmatrix}\begin{bmatrix}A_3\\A'_3\end{bmatrix}
\end{equation}
Setting $z=L$:
\begin{equation} \label{eq:boundaryzL}
\begin{bmatrix}e^{iqL}&e^{-iqL}\\iq\,e^{iqL}&-iq\,e^{-iqL}\end{bmatrix}\begin{bmatrix}A_2\\A'_2\end{bmatrix}=\begin{bmatrix}e^{ikL}&e^{-ikL}\\ik\,e^{ikL}&-ik\,e^{-ikL}\end{bmatrix}\begin{bmatrix}A_3\\A'_3\end{bmatrix}
\end{equation}
Using the expressions in Eqs. \eqref{eq:boundaryz0} and \eqref{eq:boundaryzL}, we can eliminate the $A_2$ and $A'_2$ terms and solve for $A_3$ and $A'_3$ in terms of the QM transfer matrix and $A_1$ and $A'_1$:
\begin{widetext}
\begin{equation} \label{eq:TmatrixQM} 
\begin{bmatrix}A_3\\A'_3\end{bmatrix}= \\
    \frac{e^{-ikL}}{2kq}\,\begin{bmatrix}2kq\cos{(qL)}+i(q^2+k^2)\sin{(qL)} & i(q^2-k^2)\sin{(qL)}\\e^{2ikL}(-i(q^2-k^2)\sin{(qL)}) & e^{2ikL}(2kq\cos{(qL)}-i(q^2+k^2)\sin{(qL)})\end{bmatrix}\begin{bmatrix}A_1\\A'_1\end{bmatrix}
\end{equation}
\end{widetext}

Once past the barrier, the particle is  again propagating freely towards $+\infty$ with no further reflection, so we can set $A'_3=0$:
\begin{equation} \label{eq:refl}
\begin{split}
    A'_3 = 0 &= \frac{e^{ikL}}{2kq}(-i(q^2-k^2)\sin{(qL)})A_1 \\ &+ \frac{e^{ikL}}{2kq}(2kq\cos{(qL)}-i(q^2+k^2)\sin{(qL)}) A'_1
\end{split}
\end{equation}
Similarly, we can solve for $A_3$ in terms of $A'_1$ and $A_1$:
\begin{equation}
\begin{split}
    A_3 &= \frac{e^{-ikL}}{2kq}(2kq\cos{(qL)} +i(q^2+k^2)\sin{(qL)}) A_1 \\ &+ \frac{e^{-ikL}}{2kq}(i(q^2-k^2)\sin{(qL)}) A'_1
\end{split}
\end{equation}
Finally, we can solve for the complex reflection and transmission coefficients, $r$ and $t$, which are analogous to the S-parameters $S_{11}$ and $S_{21}$ in EM:
\begin{equation} \label{eq:rQMabove}
r = \Big|\frac{A'_1}{A_1}\Big| = \frac{i(\frac{q}{k}-\frac{k}{q})\sin(qL)}{2\cos(qL)-i(\frac{q}{k}+\frac{k}{q})\sin(qL)}
\end{equation}

\begin{equation} \label{eq:tQMabove}
t = \Big|\frac{A_3}{A_1}\Big| = \frac{2e^{-ikL}}{2\cos(qL)-i(\frac{q}{k}+\frac{k}{q})\sin(qL)}
\end{equation}
Simple inspection reveals that Eqs. \eqref{eq:rQMabove} and \eqref{eq:tQMabove} are identical in form to the analogous EM S-parameter expressions in Eqs. \eqref{eq:S11pos} and \eqref{eq:S21pos}.

The case above was devived for above-barrier propagation.  For below barrier tunneling, the wavefunction becomes evanescent, the wavenumber $q\rightarrow i\rho$, and the expressions for the complex reflection and transmission coefficients become:
\begin{equation} \label{eq:rQMbelow}
r = \frac{-(\frac{\rho}{k}-\frac{k}{\rho})\sinh(\rho L)}{2\cosh(\rho L) +(\frac{\rho}{k} +\frac{k}{\rho})\sinh(\rho L)}
\end{equation}
\begin{equation} \label{eq:tQMbelow}
t = \frac{2e^{-ikL}}{2\cosh(\rho L) +(\frac{\rho}{k} +\frac{k}{\rho})\sinh(\rho L)}
\end{equation}

\section{Electromagnetic Barrier}

 Several researchers have previously investigated EM wave tunnelling in different 
systems; low-dielectric-constant region separating two regions of higher dielectric constant \cite{zhu},  waveguides (below cutoff),  naturally occurring single negative materials ($\epsilon<0$ and  $\mu>0$) \cite{tai,Dragila},  and more recently in single negative artificial electromagnetic media, see Liu \cite{Liu}, and references 3-19 therein. 
The majority of these studies focused on the analysis of an EM wave tunnelling in its own right and not as an analogy for QM tunnelling. 
 
 In this paper we focus on using an EM system as an analog for the 1D quantum barrier system shown in figure \ref{fig:pb}, two semi-infinite leads either side of a barrier region of width $d$.  
 Key to developing EM analogs of QM tunnelling is understanding wave propagation across interfaces. The QM wavefunction $\psi$ considered is a scalar, whereas in general electromagnetic waves are vectors. If we consider the case of a propagating EM wave incident upon an interface between two media, where the electric field is parallel at all points in space to the interface (i.e. a transverse electric (TE) wave), then the equivalent QM wavefunction $\psi$ of an electron incident on an interface between two media becomes directly analogous to the electric field component (the tangential component of the EM field)\cite{hend}. As discussed by Dragoman \cite{Drago}, this is consistent with the continuity condition of the wavefunction and the tangential component of the TE wave at the interface. The other boundary condition for an electron wave at an interface between two media is the continuity of $\nabla \psi \cdot \hat{z}$,  which corresponds to the continuity of the non-tangential magnetic field component of the EM wave \cite{Drago}.
 
To produce the EM equivalent of the QM rectangular potential barrier/well, shown in figure \ref{fig:pb}, requires a system where the EM energy, given by $E=\hbar \omega$, fulfils the condition $E<V_b$, resulting in an evanescent wave with an imaginary propagation constant $\beta$. Conversely for $E>V_b$ the propagation constant $\beta$ is real and the wave propagates. Eq. \eqref{eq:eqnrs3} indicates that $E<V_b$ ($E>V_b$) occurs when the media of region II (in figure \ref{fig:pb}) has constitutive parameters of the form permittivity $\epsilon<0$ ($\epsilon>0$) and  permeability $\mu>0$. For example, consider a TE wave propagating in $\hat{z}$ where the electric field is given by $E_0\exp[i(k z - \omega t)]\hat{x}$, in a medium with $k= \omega \sqrt{\mu \epsilon}$, for $\mu >0$ and $\epsilon < 0$ we see that the EM wave is evanescent. 
 A medium that produces this response can be realised by using an artificial dielectric constructed from a 2D wire array media, with a frequency dependent permittivity and a constant positive permeability\cite{brown}.
 Although the permittivity is a frequency dependent complex quantity, where the imaginary component relates to loss/absorption in the material, accounting for loss/absorption in QM calculations is difficult (see for example the paper by Dung\cite{los}), to avoid this issue (as discussed later) we only consider artificial dielectrics designed with negligible loss.

 \begin{figure}
\begin{center}
\includegraphics[width=0.7\columnwidth]{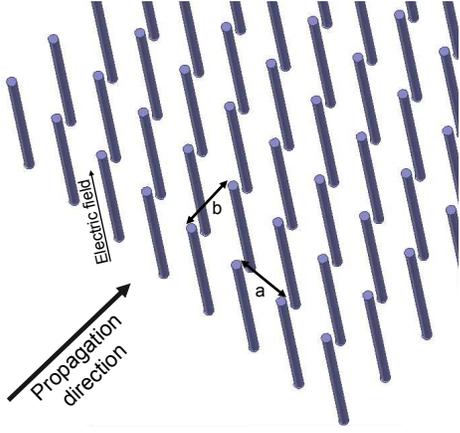}
\caption{\label{fig:wirearr} Schematic of a periodic wire array media, of infinitely long wires spaced $a$ and $b$ apart where the electric field of the incident wave is polarised parallel to the wires of the medium\cite{brown}.}
\end{center}
\end{figure}

\subsection{Wirearray Media}
Using the approach of Brown \cite{brown} the refractive index $n$ for an artificial dielectric made from infinite long wires of radius $r$, separated by a distance $b$ in the direction of propagation of normal incidence EM wave, and a distance $a$ perpendicular to wave propagation (shown in figure \ref{fig:wirearr}) is given by\cite{brown}, 
\begin{equation} \label{eq:eqnrs4}
n = \frac{\lambda_0}{2 \pi b} \arccos{ \left[ \cos{(2 \pi b / \lambda_0)}
 + \frac{\lambda_0}{2 a} \frac{\sin{( 2 \pi b/\lambda_0)}} {\ln{(a/2 \pi r)}}
\right]}
\end{equation}
where $\lambda_0$ is the free-space wavelength of the incident wave, and the impedance $Z$ of the system is given by\cite{brown},
 \begin{equation} \label{eq:eqnrs4b}
Z_b = Z_0 \frac{\tan(\pi b/ \lambda_0)}
{\tan(\pi b n / \lambda_0)},
\end{equation}
 with the impedance of free space $Z_0 = \mu_0 / \epsilon_0 = 376.8 \Omega$ and the permittivity ($\epsilon$)  determined directly from $\epsilon = n / Z$. 
 \begin{figure}
\begin{center}
\includegraphics[width=0.95\columnwidth]{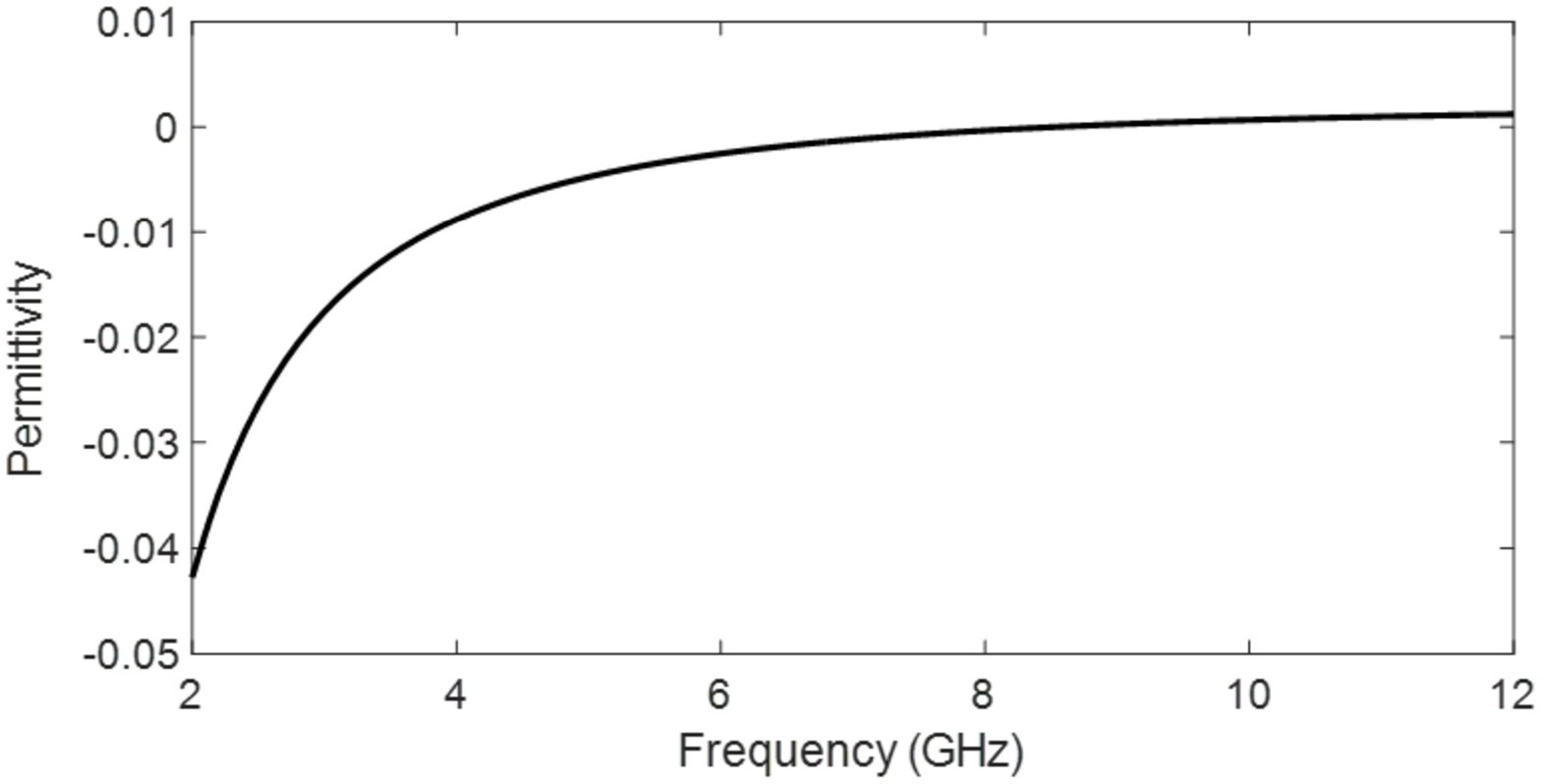}
\caption{\label{fig:br-per} Permittivity calculated using Eq. \eqref{eq:eqnrs4} with r=0.04mm a=10mm b=5mm.}
\end{center}
\end{figure}
 Figure \ref{fig:br-per} shows the permittivity of the wire array media shown in figure \ref{fig:wirearr} calculated using Eq. \eqref{eq:eqnrs4} with $r=0.04$ mm, $a=10$ mm and $b=5$ mm. At low frequency the permittivity is highly negative, approaching zero as frequency increases and becomes positive at $8.5$ GHz.
 
\begin{figure}
\begin{center}
\includegraphics[width=0.9\columnwidth]{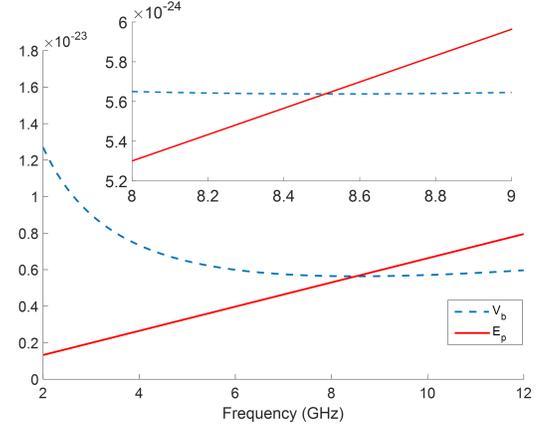}
\caption{\label{fig:br-qm} Equivalent quantum potential and particle energy calculated using data from figure \ref{fig:br-per}. Inset shows reduced frequency range of main graph.}
\end{center}
\end{figure}

To examine the correspondence between QM and EM we rearrange Eq. \eqref{eq:eqnrs3} to give an expression for the effective barrier height $V_b$ in terms of the refractive index 
,
\begin{equation} \label{eq:barrierEM}
    V_b = E_p - \frac{h^2 f^2 n^2}{2m_{eff} c_0^2},
\end{equation}
where $E_p$ the equivalent energy of the photon is given by,  
\begin{equation} \label{eq:E0EM}
    E_p = hf,
\end{equation}
 $f$ the frequency of the incident electromagnetic wave
 and the effective photon mass $m_{eff}$ is given by, 
\begin{equation} \label{eq:photonmass}
    m_{eff} = \frac{hf}{c_0^2}.
\end{equation}
This enables us to calculate the equivalent quantum effective barrier $V_b$ (shown in fig \ref{fig:br-qm}), using the data from figure \ref{fig:wirearr} to determine $n$.
 $E_p$ the energy behaves as expected increasing with frequency, whereas the potential demonstrates a variation with frequency due to the frequency dependence of the permittivity. However, as shown in the insert graph of figure \ref{fig:br-qm}, in the vicinity of the crossing point of $E_p$ and $V_b$, the variation of $V_b$ with respect to frequency is negligible.  As there are regions within this frequency range where $E_p < V_b$ and $E_p > V_b$, the wire array medium used in the EM system is an appropriate analog for a QM barrier where $E_p$ can be above or below the barrier potential $V_b$. 
 
%
 
 \subsection{Simulation of wire array media}
 The approach of Brown\cite{brown}
  uses several assumptions;  the wire array is semi-infinite in the direction of wave propagation,  infinite perpendicular to the direction of propagation, with no loss. To use this medium as an analog for a QM barrier of a finite width, a wire array medium of finite thickness in the direction of wave propagation is modeled using the commercial finite element method 3D EM solver HFSS from Ansys, which it has been successfully validated with close agreement against a number of experimental measurements for various EM systems.

 Following the approach of Brown we model wave propagation in the z-direction across a single array of $N$ Cu wires (radius $r$).  The wire array medium is embedded in a vacuum box with lattice spacing $a$ in the x-y plane, with Bloch-Floquet boundary conditions to model an infinite block of material in the x-y directions.  The material has finite thickness in the z-direction which is equal to the lattice spacing in the z-direction, $b$ multiplied by the number $N$ of rows of Cu wires.  The vacuum box is at least 10 wavelengths in the z-direction, with ports at either end to emit and absorb TE waves oriented such that the electric field is parallel to the wires, producing a system that mimics a QM potential barrier $V_b$ of thickness $N b$ with infinitely long leads either side (see figure \ref{fig:wirearr}). These EM simulations naturally yield the S-parameters describing EM wave scattering from the system that can be directly compared with the equivalent QM S-parameters discussed in the previous section.

For the EM case the measured S-parameters can be used 
 to determine the refractive index, permittivity and permeability of our artificial dielectric wire array media
using a modified  Nicolson-Ross-Weir (NRW) technique \cite{nic,weir} adapted  to account for possible negative responses in the real components of the refractive index and permittivity/permeability\cite{smith-ext}. The NRW uses a closed-form expression allowing the complex form of the refractive index to be determined directly from S-parameter measurements, and is also robust to experimental error\cite{nic,weir}.
This approach expresses the impedance $Z=Z' +i Z''$ and refractive index   $n=n'+n''$ of the material as\cite{smith-ext};

\begin{align} 
\label{eq:ex3}
Z &= \pm \left [ \frac{(1+S_{11})^2 - S_{21}^2}{(1-S_{11})^2 - S_{21}^2}  \right ]^{1/2} \\
\label{eq:ex4}
n' &=  \pm \frac{1}{kd}  \Re \left [  \cos^{-1}  \left ( \frac{1- S_{11}^2 +S_{21}^2 }{2S_{21}}\right ) \right ] + \frac{2 \pi m}{k d} \\
\label{eq:ex5}
n'' &=  \pm \frac{1}{kd} \Im \left  [  \cos^{-1}  \left ( \frac{1- S_{11}^2 +S_{21}^2 }{2S_{21}}\right ) \right ]
\end{align}
where $k_0 = 2 \pi / \lambda_0$ is the free space wave vector and $m$ an integer for the correct solution branch, chosen to ensure that $n'$ is continuous across the frequency range. From this the effective permittivity and effective permeability can be determined directly from $\epsilon_{f} = n / Z$ and   $\mu_{f}= nZ$. Note in this case $n$,$Z$, $\epsilon_f$ and $\mu_f$ are complex, where the imaginary component of $\epsilon_f$ and $\mu_f$ describe loss.

\begin{figure}
\begin{center}
\includegraphics[width=1\columnwidth]{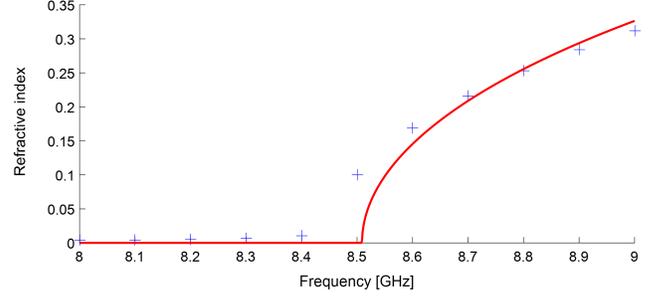}
\caption{\label{fig:simcomp} Comparison of the refractive index calculated by simulation (Blue markers) and theory (Red line).}
\end{center}
\end{figure}

Figure \ref{fig:simcomp} presents a comparison of the refractive index calculated from Eq. \eqref{eq:ex4} (cross markers) using S-parameters calculated from an HFSS simulation for a wire array consisting of a 9-wire deep barrier ($r=0.04mm$, $a=10mm$, $b=5mm$) and the theoretical approach calculated by Eq. \eqref{eq:eqnrs4} (red line). A key difference between the simulated and theoretical  calculations is the simulated system incorporates loss from the Cu wires, while the theoretical system assumes lossless, perfectly conducting wires.  This loss is generally found to be very small and the agreement between the theoretical and simulated refractive index is within a $1\%$ variation over most of the frequency range. Larger differences between the simulated and theoretical refractive indices occur at the inflection point (i.e at the point permittivity, see figure \ref{fig:br-qm} becomes zero), where the equivalent QM particle energy transitions from below to above barrier energy in figure \ref{fig:br-qm}.  Despite the small loss term, the simulation results show the imaginary component of $\epsilon$ (relating to loss) are at least two orders of magnitude smaller than the real component of $\epsilon$ and furthermore, the permeability $\mu$ in the simulation results maintains a constant positive value across the simulated frequency range.

\begin{figure}
\begin{center}
\includegraphics[width=1\columnwidth]{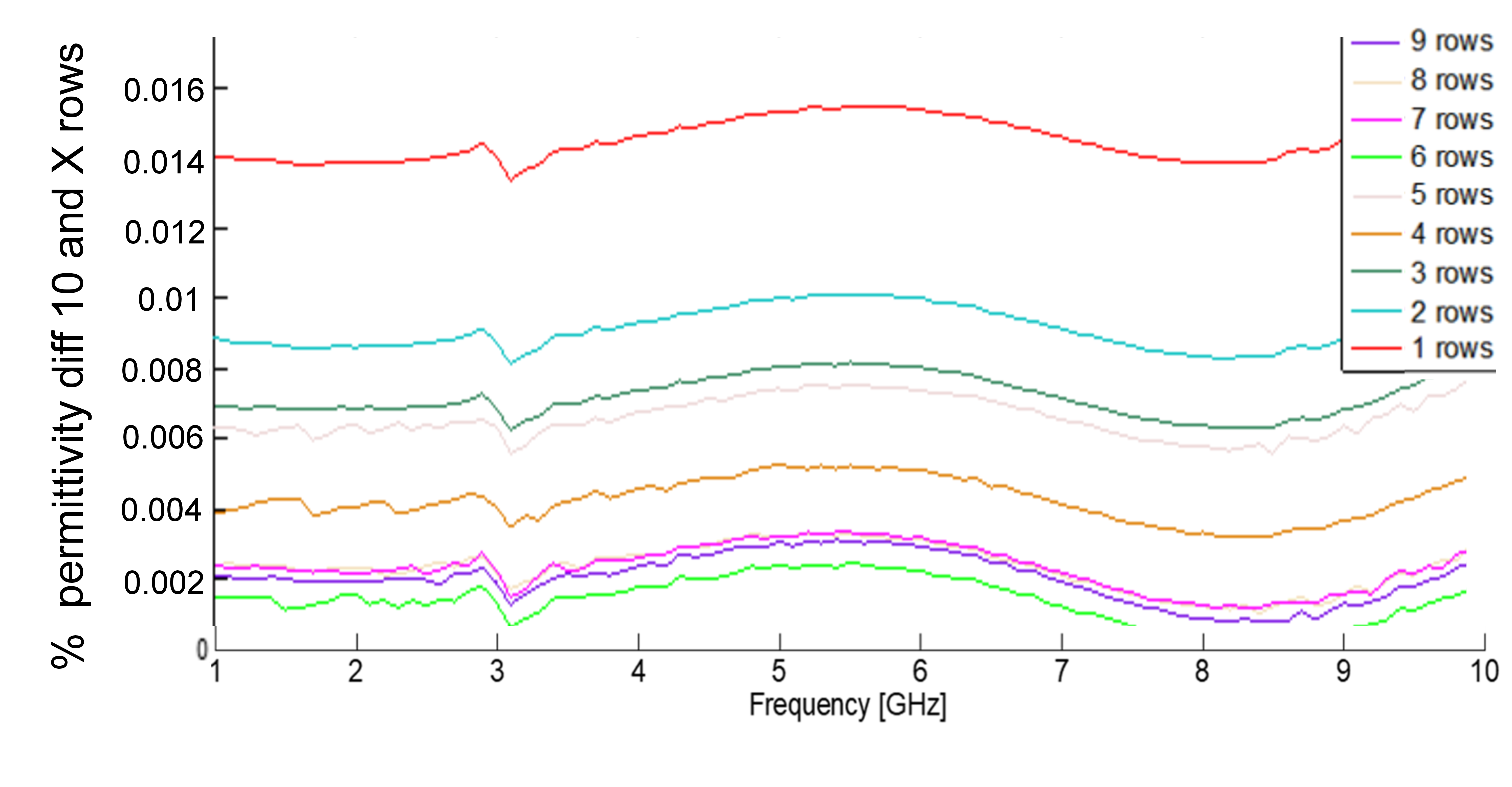}
\caption{\label{fig:perdif} Percentage difference in permittivity for wire arrays with X rows compared to a media consisting of 10 rows of wires.}
\end{center}
\end{figure}

In terms of assessing  impact on the QM barrier potential $V_b$ from varying lengths $N b$ of the wire media we simulated systems with a varying number of wires $N$ in the z-direction. From the simulated S-parameters we calculated the permittivity of media consisting of between $1 -10$ wires in the direction of propagation and calculated the variation in permittivity from the 10 wire row system, at each frequency point. The results from this calculation are shown in Figure \ref{fig:perdif}, where we see for systems consisting of more than 4 rows of wires the variation of the permittivity, and hence variation in $V_b$, is less than $0.04\%$ and less than $0.02\%$ between 8 and 9 GHz the region of our constant $V_b$. Hence, wire  media, of finite length can be used to form analogs of QM barriers of finite and varying length, by varying the number of wires in the barrier region without substantially altering $V_b$.  
 
\section{Results and Discussion}

In this section, we consider two simple textbook examples of an analytical plane wave incident on a simple rectangular barrier and on a delta function barrier.  For the rectangular barrier, we will show how the QM parameters and reflection and transmission coefficients are extracted from an equivalent EM system, as well as how an effective barrier potential energy can be calculated for the EM example.  Next, we show how the rectangular barrier approaches the delta function limit for both systems. Finally, we will show the effects of varying height and width independently for each system.

\subsection{Rectangular Barrier }

Beginning with the free space QM wave vectors, we calculate the corresponding free space energies:
\begin{equation} \label{eq:E0QM}
    E_0 = \frac{\hbar^2 k^2}{2m_e}
\end{equation}
and frequencies:
\begin{equation} \label{eq:nuQM}
    \nu = \frac{c_0 k}{2\pi}
\end{equation}
where $m_e$ is the rest mass of the electron and $c_0$ is the speed of light in a vacuum. By rearranging Eq \eqref{eq:eqnrs3} and mapping the refractive index to the QM frequency range we can solve for the barrier height, $V_b$:
\begin{equation} \label{VQM}
V_b = E_0 - h^2\nu^2n^2/(2m_e c_0^2)
\end{equation}
where $E_0$ is calculated in Eq. \eqref{eq:E0QM},  $\nu$ is the QM electron frequency (Eq. \eqref{eq:nuQM}) and $n$ is  calculated from the EM S-parameters which are generated from HFSS simulations. The EM system consists of a 5 wire deep wire lattice with $a=10$ mm, $b=5$ mm, $r=0.04$ mm, using Bloch-Floquet boundary conditions to mimic an infinite sheet of thickness $N b$.  Although there is an imaginary component to the barrier potential $V_b$ due to loss in the Cu wires that gives rise to a complex $\epsilon$, this is negligible ($\approx 10^{-5}$) in comparison to the real component, and we consider only the real component of $V_b$.  Because $V_b$ is calculated from a frequency-dependent refractive index, it too will have a frequency dependence.  This is contrary to how we usually think of a quantum mechanical energy barrier, which is typically constant with respect to frequency.  As discussed in the previous section, however, if we restrict the EM frequency range (QM equivalent $E_p$), we can then consider $V_b$ to be constant for changing $E_p$.

Similarly, the wavenumber in the barrier region, $q$, is calculated at each frequency:
\begin{equation} \label{eq:qbQM}
    q = k\sqrt{\frac{(V_b - E_0)}{E_0}}
\end{equation}
and will vary both as a function of $E_0$ and $V_b$, rather than just $E_0$.

{The barrier width $L$ also has a frequency dependence, as it scales as both a function of refractive index $n$ and the EM free space wavenumber:
\begin{equation} \label{eq:lengthQM}
    L = \frac{n k_{EM} d}{q}
\end{equation}
where $k_{EM}$ is the wavenumber of the corresponding electromagnetic frequency, $d$ is the thickness of the dielectric material, and $q$ is the QM wavenumber in the barrier region.}

The frequency dependence of the QM potential barrier, and the dependence of the barrier thickness and wavenumber on the quantum barrier potential can all be minimized through careful choice of the EM frequency range and the parameters of the wire array medium in that range.  The frequency range from the HFSS simulation was truncated to include only frequencies at which the wire array medium acts like and effective medium, thereby minimizing this dependence.


\begin{figure}
    \centering
    \includegraphics[width=\columnwidth]{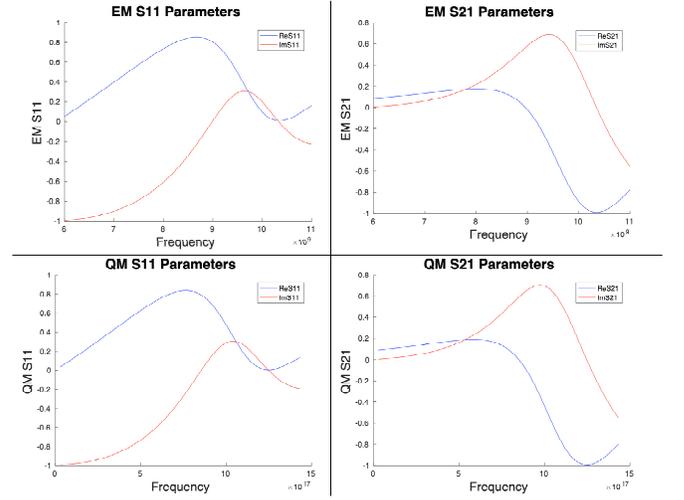} 
    \caption{A comparison of the real and imaginary parts of the  S11 and S21 coefficients from the EM HFSS simulation compared to those of the QM simulation where the parameters were derived from the EM system.  Note that while the frequencies differ for the EM and QM systems, the lineshapes and their amplitudes are identical.} 
    \label{fig:Sparams}
\end{figure}

\begin{figure}
    \centering
    \includegraphics[width=\columnwidth]{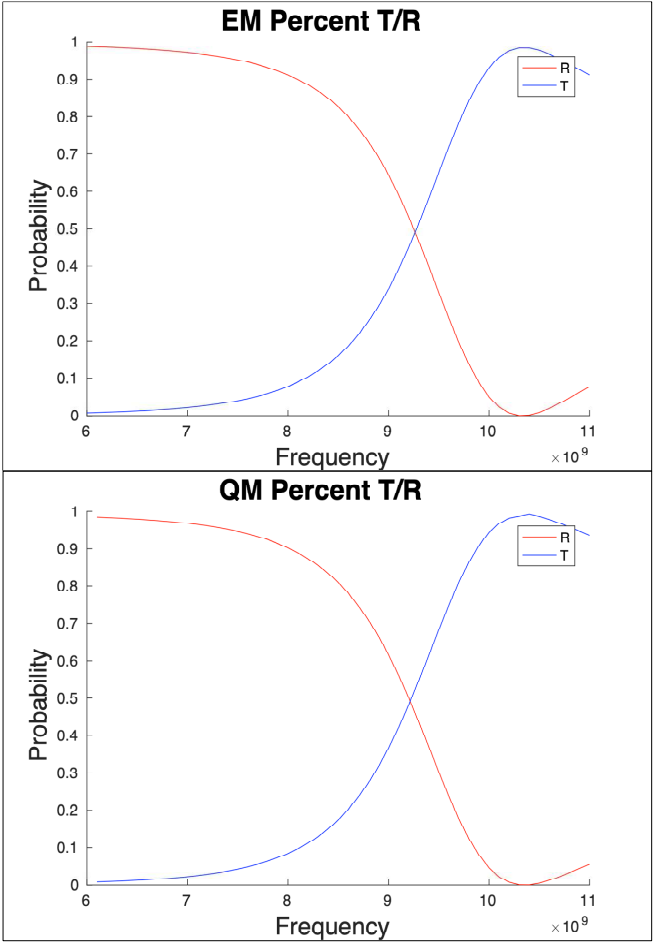} \\
    \caption{A comparison of  transmission and reflection from the EM HFSS simulation and the corresponding QM system.}
    \label{fig:percents}
\end{figure}

We use these values of $k$, $q$ and $L$ to find the complex reflection and transmission coefficients in Eqs. \eqref{eq:rQMabove} and \eqref{eq:tQMabove}.  We plot the real and imaginary components of $r$ and $t$ and compare them to the real and imaginary components of the S-parameters in Figure \ref{fig:Sparams}.  Although it is more common to plot the transmission and reflection, as shown in Figure \ref{fig:percents}.

{Next, we consider the QM/EM correspondence, by calculating the effective barrier potential ($V_b$) energy ($E_p$) of the wire array medium for the electromagnetic case from Eqs. \eqref{eq:barrierEM}, \eqref{eq:E0EM}, \eqref{eq:photonmass}. Figure \ref{fig:E0Vbvsf}  compares  $E_p$ and $V_b$ vs. frequency for EM system to the corresponding QM system.  While the lineshapes are different in the two plots as a consequence of different frequency dependencies (linear for EM, quadratic for QM), both plots show the incident free space energy to be below the barrier height at the lower end of the frequency range, and above it at the higher end.  The crossing point of the $E_p$ and $V_b$ curves also occurs about halfway through the respective frequency ranges.  The insets in Figure \ref{fig:E0Vbvsf} zoom in on the region close to the crossing point in each of the panels.
\begin{figure}
    \centering
    \includegraphics[width=\columnwidth]{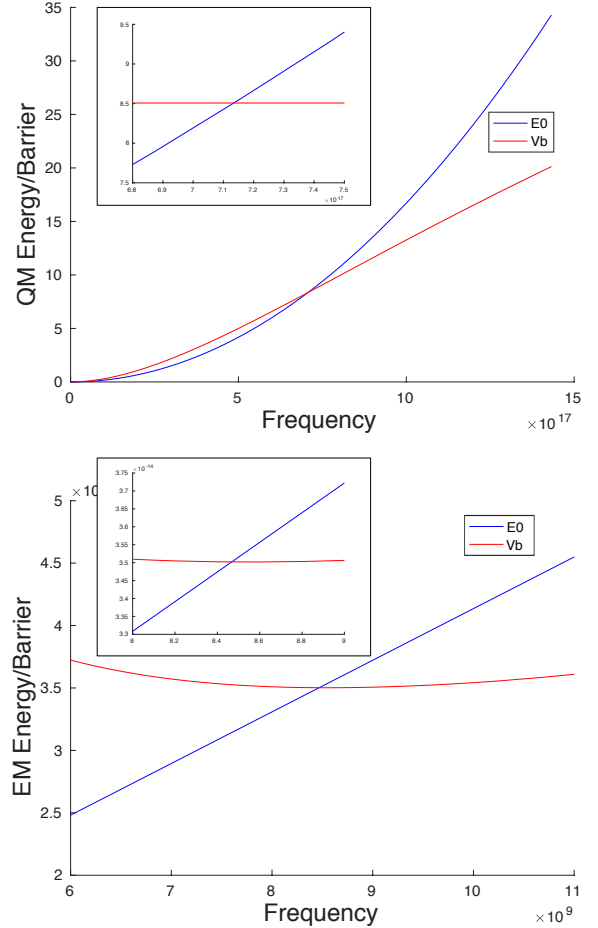}
    \caption{Incident energy and barrier height vs. frequency for the QM and EM systems, inset in each panel shows a close-up of the crossing point.}
    \label{fig:E0Vbvsf}
\end{figure}
If a single value of the barrier height is chosen close to the crossing point of the EM plot, the calculated value of $V_b$ for the QM system can be held constant here to give the result with which we are more familiar.}

\subsection{Delta Function Limit }
 The second analytical example we consider is the $\delta$-function potential.  {Although the $\delta$-function potential is a fairly well-studied analytic textbook example in quantum mechanics \cite{Cahay2017a, Jensen2019}, its derivation is included in an Appendix(??)for clarity and convenience.}  If we begin with the rectangular barrier in Figure \ref{fig:pb}, and imagine increasing its height to $\infty$ and reducing its width to zero while keeping the ``area" under the barrier constant, we end up with a $\delta$-function barrier.  
Beginning with the above barrier expressions for the t,r coefficients of a rectangular potential barrier, Eqs. \eqref{eq:rQMabove} and \eqref{eq:tQMabove}:
\begin{equation}
r = \Big|\frac{A'_1}{A_1}\Big| = \frac{i(\frac{q}{k}-\frac{k}{q})\sin(qL)}{2\cos(qL)-i(\frac{q}{k}+\frac{k}{q})\sin(qL)}
\nonumber
\end{equation}
\begin{equation}
t = \Big|\frac{A_3}{A_1}\Big| = \frac{2e^{-ikL}}{2\cos(qL)-i(\frac{q}{k}+\frac{k}{q})\sin(qL)}
\nonumber
\end{equation}
We make two assumptions.  First, the energy $E$ of the incident particle is held constant.  Secondly, the "area" under the potential barrier, or the product of barrier height $V_b$ and barrier width $L$, remains constant as $V_b \rightarrow \infty$ and $L \rightarrow 0$.  Next, we revisit the definition of barrier-region wavenumber, $q$, Eq. \eqref{eq:qbarrier}:
\begin{equation}
q=\sqrt{\frac{2m(E-V_b)}{\hbar^2}}
\nonumber
\end{equation}
Since $E$ remains fixed while $V_b \rightarrow \infty$, $\frac{k}{q} \rightarrow 0$.  Additionally, since $L \rightarrow 0$, the exponential term $e^{i k L} \rightarrow 1$.  The coefficient expressions can be rewritten as:
\begin{align*}
    r &= \frac{i \frac{q}{k}\sin(qL)}{2\cos(qL)-i \frac{q}{k}\sin(qL)} \\
    t &= \frac{2}{2\cos(qL)-i \frac{q}{k}\sin(qL)}
\end{align*}
Now we consider the behavior of $qL$ in the $\delta$-function limit.  Since the product of $V_b L$ is constant, $L \rightarrow 1/\infty$ as ${V_b} \rightarrow \infty$. As $V_b \rightarrow \infty$, $q \rightarrow \sqrt{\infty}$.  Therefore, since $L \rightarrow 0$ faster than $q \rightarrow \infty$, we can see that $qL \rightarrow 0$.  As $q$ also becomes imaginary as  $V_b \rightarrow \infty$, the functions will become hyperbolic, but the resulting series expansions are the same for both the hyperbolic and trigonometric forms, so we expand the trig functions into a Taylor series and take the small argument limit of $qL$.  The coefficients are  rewritten as:
\begin{align} 
\label{eq:rdellim}
    r &= \frac{\frac{i q^2 L}{k}}{2-\frac{i q^2 L}{k}} \\
\label{eq:tdellim}
    t &= \frac{2}{2-\frac{i q^2 L}{k}}
\end{align}
We see that in the limit where $V_b \rightarrow \infty$:
\begin{equation}
    q^2 = \frac{2m(E-V_b)}{\hbar^2} \approx -\frac{2mV_b}{\hbar^2}
\end{equation}
Substituting in this value of $q^2$ into Eqs. \eqref{eq:rdellim} and \eqref{eq:tdellim} and rearranging, we are left with
\begin{align}
r &= \frac{1}{\frac{i \hbar^2 k}{mV_bL}-1} \\
t &= \frac{1}{1-\frac{mV_bL}{i \hbar^2 k}}
\end{align}
Previously, we defined $\lambda = V_bL$ and $\gamma = \frac{k}{\lambda}$, we can rewrite the reflection and transmission coefficients as,
\begin{equation*}
r = \frac{1}{\frac{i \hbar^2 \gamma}{m}-1}; ~~t = \frac{1}{1-\frac{m}{i \hbar^2 \gamma}}
\end{equation*}
which are simply Eqs. \eqref{eq:deltar} and \eqref{eq:deltat}, the reflection and transmission coefficients of a $\delta$-function barrier. 

Although tunneling through a $\delta$-function potential barrier is a problem more readily associated with quantum mechanics, we can construct an {approximate} electromagnetic analog.  To simulate the EM analog of the delta using HFSS the approach of Brown, detailed in the previous section, was utilised. The initial conditions for the wire array were,($N$) 5 rows of wires, $r=40 \mu$m, $a=10$ mm, $b=10$ mm simulated at a frequency $f=3$ GHz. The simulations then proceed to reduce $b$ in steps of $1$ mm.
In the electromagnetic simulation, the total thickness of the wire array medium is $L$ (a product of the lattice constant $b$ and the number of rows of wires $N$ ), while the effective barrier height is given by Eq. \eqref{eq:barrierEM}. 
As $E_0$ the energy of the incident radiation is defined by $f$ and $m_{eff}$, are constant the only quantity that can be varied to a adjust the height of the barrier is $n=\sqrt{\epsilon \mu}$. Hence to keep the product $V_b L$ constant, for analogy to the quantum mechanical example above, as $b$ is reduced we change the lattice constant $a$ in Eq. \eqref{eq:eqnrs4} to increase $V_b$.
Table \ref{tab:One} lists the parameters for $b$ and $a$ used in the simulations.
\begin{table}
\caption{\label{tab:One} Lattice parameters $a$ and $b$ ($mm$) for constant $V_b b$.}
\begin{center}
\begin{tabular}{c| c c c c c c c c c c}
{ $b$}& {\small 10 }& {\small 9} & {\small 8} & {\small 7} & 
{\small 6} & {\small 5} & {\small 4} & {\small 3} & {\small 2} & {\small 1}\\
{ $a$}& {\small  10 }& 
{\small 9.876}&
{\small9.763}&
{\small9.660}&
{\small9.567}&
{\small9.483}&
{\small9.409}&
{\small9.343}&
{\small9.285}&
{\small9.237}
\\
\end{tabular}
\end{center}
\end{table}

\begin{figure}
    \centering
    \includegraphics[width=\columnwidth]{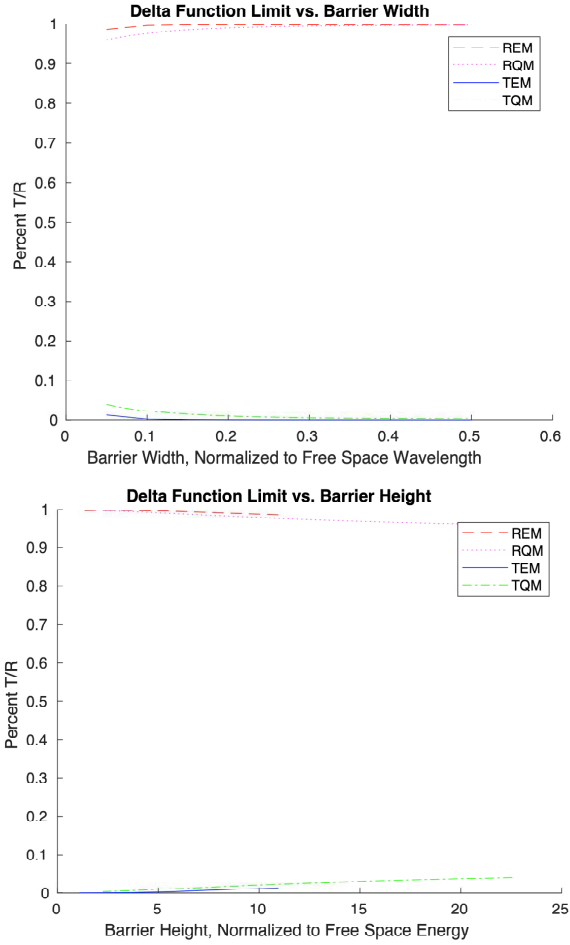}
    \caption{The $\delta$-function limit of a  quantum mechanical rectangular barrier.  Transmission and reflection probabilities are shown vs. barrier thickness and barrier height.}
    \label{fig:delta}
\end{figure}

 Using the lattice parameters given in table \ref{tab:One} HFSS simulations were done to determine the reflection and transmission coefficients as $b$ is decreased.   The smallest length we can achieve for EM with this configuration is $b=1mm$.  Figure \ref{fig:delta} shows the change in the transmission and reflection coefficients for both systems plotted together as a function of barrier width and barrier height.  The small divergence in behavior between the QM and EM systems as the delta function limit is approached can be attributed to the limitations of HFSS.  In addition to the simulation runtime becoming prohibitively long as the lattice spacing in the wire array medium is decreased, in this limit HFSS also begins to interpret the medium as a solid sheet of metal, which would lead to overestimation of the reflection coefficient and underestimation of the transmission coefficient, as seen in Figure \ref{fig:delta}.  {For these reasons, the EM analog can only approximate the $\delta$-function, not reproduce it exactly, and as the lattice spacing reduces the barrier width below $0.05 \lambda$ the EM analog breaks down.}  Nonetheless, the same trends are evident in both systems.  For the QM system compared to barrier height, the simulation was carried out closer to the $\delta$-function limit so that the asymptotic behavior of the transmission and reflection traces can be seen more clearly.  These values approach the theoretical delta function limit predicted by the absolute squares of Eqs. \eqref{eq:deltar} and \eqref{eq:deltat}.

\subsection{Parameter variation}

{Now that we have illustrated the correspondence between EM and QM for a couple of simple, analytical examples, we will show how the parameters of height and width can be varied independently in the EM system by adjusting the parameters of the wire array medium using Brown's approach.}

\begin{figure}
    \centering
    \includegraphics[width=0.8\columnwidth]{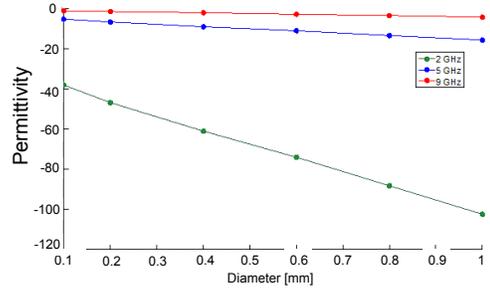}
    \caption{Variation of the permittivity vs diameter for a wire array media, plotted for three frequencies 2, 5, 9 GHz. }
    \label{fig:epsdia}
\end{figure}

{For the analog EM simulations of a QM barrier, consisting of a wire array, to vary the barrier potential ``height" ($V_b$), whilst keeping the barrier width a constant, we vary the diameter of the wires of the media. Increasing the wire diameter increases the permittivity (Eq. \eqref{eq:eqnrs4b}), which increases $V_b$ (Eq. \eqref{eq:barrierEM}). Figure \ref{fig:epsdia} shows the HFSS simulation results for variation of permittivity vs diameter of the wire for three frequencies 2, 5, 9 GHz, all other parameters of the array were held constant, as expected the magnitude of the permittivity increases with increasing wire diameter. As seen, the change in permittivity is linear with wire diameter. }

\begin{figure}
    \centering
    \includegraphics[width=0.8\columnwidth]{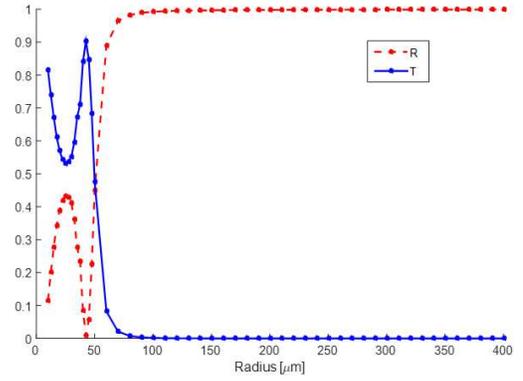}
        \caption{Transmission (T) and Reflection (R) vs wire radius of the wire media.}
    \label{fig:EMBarr1}
 \end{figure}

\begin{figure}
    \centering
    \includegraphics[width=0.8\columnwidth]{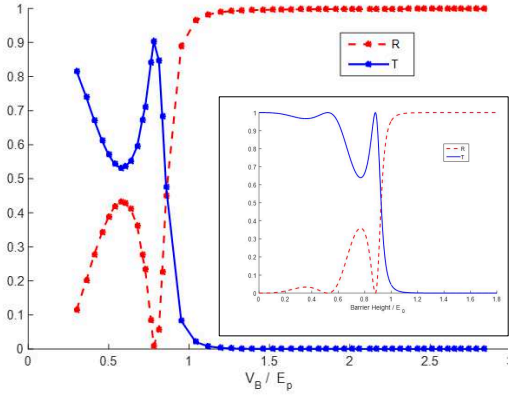}
    \caption{Transmission (T) and reflection (R) vs the  effective $V_b$ height for the equivalent EM system of varying wire media radius; inset shows the analogous QM system.}
    \label{fig:EMBarr2}
\end{figure}

{Using this approach of varying the wire diameter for a fixed frequency of $9$ GHz to vary the permittivity in a controlled linear manner, we can determine a QM $V_b$ equivalent of the system from the EM simulation results of Transmission (T) and Reflection (R), using Eq. \eqref{eq:barrierEM}.  Figure \ref{fig:EMBarr1} shows $T$ and $R$ verses varying wire media radius. Figure \ref{fig:EMBarr2} shows $T$ and $R$ verses the the effective $V_b$ height for the equivalent EM system of the varying wire media radius from figure \ref{fig:EMBarr1}, where the effective $V_b$ height is normalised to the equivalent energy of the $9GHz$ input signal. The behaviour of both figures \ref{fig:EMBarr1} and \ref{fig:EMBarr2} is as expected, below a certain radius the effective permittivity is positive allowing wave propagation.  In the equivalent QM system, shown in the inset of Figure \ref{fig:EMBarr2},this is the case when $E_0>V_b$.  The oscillations in $T$ and $R$ are due to  interference between the transmitted (forward) and reflected (backward) wave from the interfaces of the barrier.}

\begin{figure}
    \centering
    \includegraphics[width=\columnwidth]{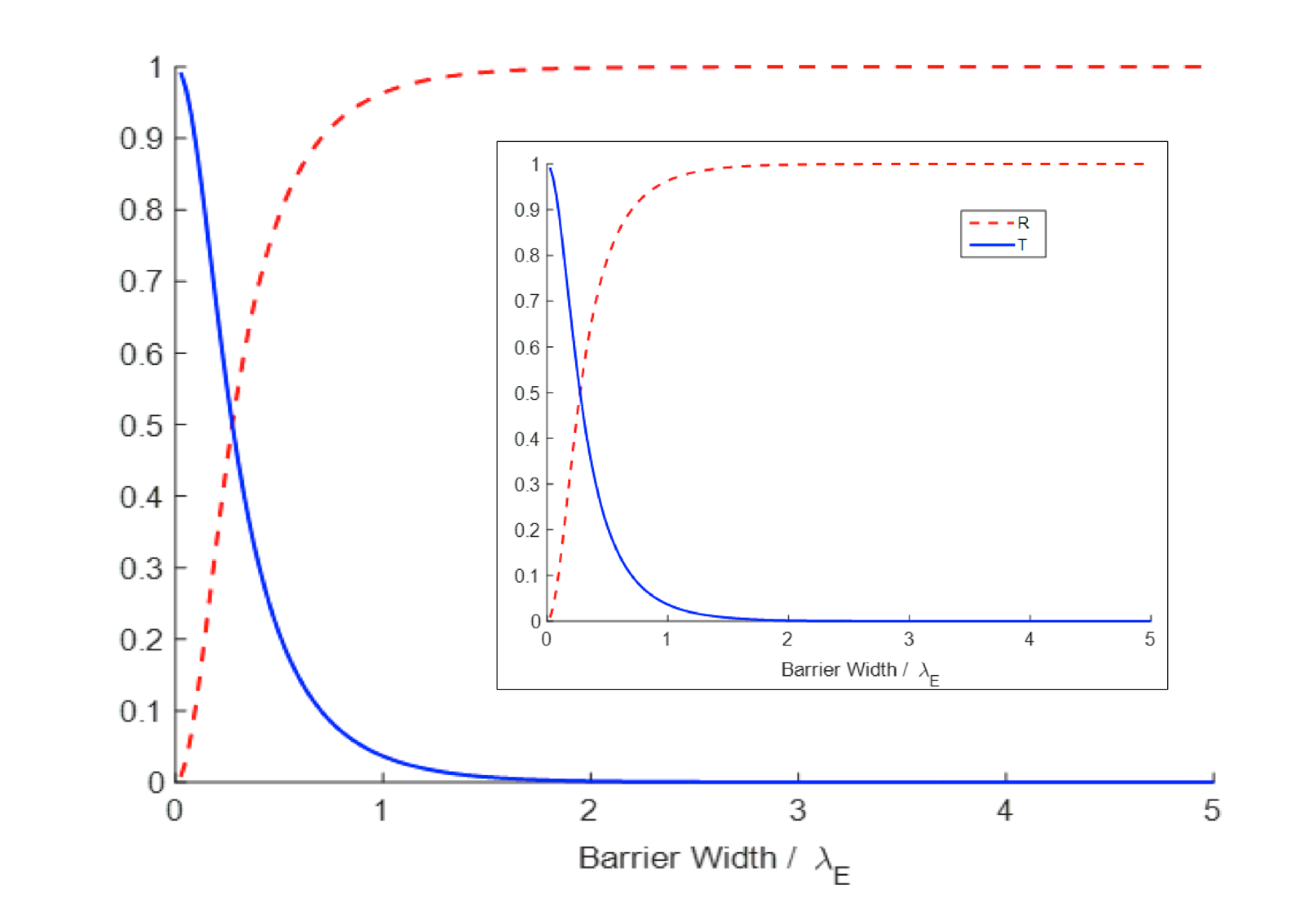}
    \caption{Variation of Transmission (T) and Reflection (R) for for a wave with energy $E_p = 0.95$ of the barrier potential $V_b$ vs the thickness of the barrier, where the barrier thickness is shown normalised to the free space wavelength of the incident wave.  The inset shows the analogous result in QM.}
    \label{fig:TR95}
\end{figure}

{Next we study the case for $E_p < V_b$, focusing on point $E_p = 0.95 V_b$. Using the EM wire array simulations where we simulate multiple width barriers  by increasing the number of wire rows $N$ between simulations. The simulated transmission and reflection coefficients  for $E_p = 0.95 V_b$ vs  barrier width are shown in Figure \ref{fig:TR95}.  This is seen to be nearly identical to the QM case (inset to Figure \ref{fig:TR95}) when the barrier width in each case is normalised to the respective incident radiation wavelength.  As expected, the transmission probability decays exponentially as the width of the barrier increases.}

\begin{figure}
    \centering
    \includegraphics[width=0.8\columnwidth]{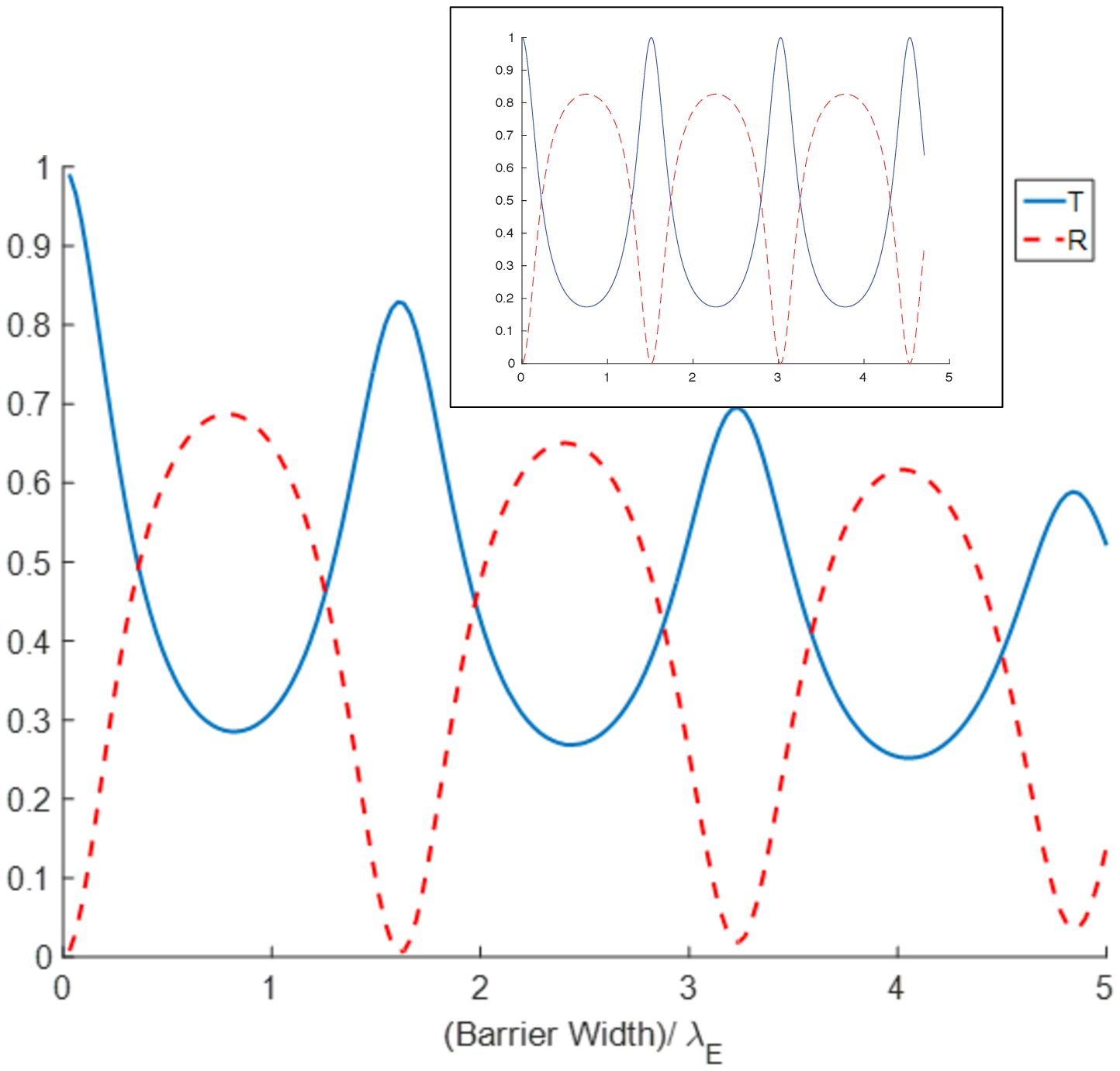}
    \caption{Variation of Transmission (T) and Reflection (R) for for a wave with energy $E_p = 1.05$ of the barrier potential $V_b$ vs the thickness of the barrier, where the barrier thickness is shown normalised to the free space wavelength of the incident wave, in the inset, the correponding QM example is shown.}
    \label{fig:TR105}
    \end{figure}

{Finally we study the case for $E_p > V_b$, focusing on point $E_p = 1.05 V_b$. Again, the thickness of the wire array medium is controlled by varying the number of wire rows, $N$.  The simulated transmission and reflection coefficients  for $E_p = 1.05 V_b$ vs  barrier width are shown in Figure \ref{fig:TR105}, with the analogous QM system shown in the inset.  As with the below barrier example, the barrier width in each case is normalized to the respective incident wavelength.  As predicted in Eq. \eqref{eq:rQMabove}, the value of the reflection and transmission coefficients oscillates with a period equivalent to the de Broglie wavelength in the material.  It is worth noting that at the energy chosen, the transmission coefficient is sometimes less than the reflection coefficient for both the QM and EM examples, even though the incident energy is greater than the barrier height.  When the product of $qL$ is equivalent to integer multiples of $\pi$, the reflection coefficient goes to zero.  As the incident energy is increased with respect to the barrier height, the period of these oscillations becomes shorter, and when the incident energy of the particle is far enough above the barrier height, the transmission coefficient is always greater than the reflection coefficient.  Another notable feature in Figure \ref{fig:TR105} is the decaying amplitude of the transmission coefficient as the width of the barrier increases in the EM example, and effect which is not observed in the analogous QM system.  This decrease is a consequence of Ohmic losses in the Cu wires that make up the wire array, which becomes more significant as more rows of wires are added to increase the thickness, thereby resulting in a decrease in the peak amplitude of the transmission coefficient with an increase in barrier thickness. This is an important effect that must be considered when designing EM analogs of QM systems, and care must be taken to ensure EM systems are designed such that the associated EM losses are small so as not to impact analog comparisons.}

\section{Conclusions}

{In this paper, we have used simple, analytical examples to demonstrate the correspondence between quantum mechanics and electromagnetism with the intent of establishing a framework for the verification and validation of our previously developed atomistic emission models \cite{Jensen2019, Jensen2019a, Jensen2021, Jensen2021a, Jensen2022}.  Beginning with the frequency-dependent EM S-parameters which can either be calculated using HFSS or obtained through laboratory measurement, the frequency-dependent refractive index vector of a wire array medium was calculated.  This refractive index vector was then mapped to a QM frequency range, enabling the calculation of the potential barrier, the wavenumber in the QM barrier region, and the barrier thickness.  These quantities were used to calculate the complex QM transmission and reflection coefficients.  When plotted against the QM frequency range, these complex coefficients were identical in lineshape to the original EM S-parameters plotted against the EM frequency range, proving the validity of the approach.  An effective EM barrier energy can also be calculated using a frequency-dependent effective photon mass.  We then looked at the delta function limit of a rectangular barrier in QM and showed the EM analog of it.  Finally, we showed how the barrier width and effective barrier height for the analogous EM system can be adjusted through variations of the wire array medium parameters, according to Brown's methodology.}

The barrier systems presented in this paper are simple examples that can be solved in both QM and EM analytically, where we show these analytic results for these simple barriers are self-consistent with themselves and  with EM numerical simulations. Demonstrating the equivalence of QM-EM calculations to numerical EM simulations, and we also discuss the limits of EM analogs of QM systems due to loss.  We also show, in figure 6, if we have more that 4 rows of wires the system behaves as a bulk media, this paths the foundations for future work where we will consider barrier systems with a spatially varying potential. For example we could consider a system of 20 rows of wire where each 4 rows of wires has a different a, b, and/or r  compared to the adjacent block of 4 rows of wires, corresponding to a system of '5' blocks of dielectric $\epsilon_1,\epsilon_2,\epsilon_3, \epsilon_4,\epsilon_5$, where each block consists of 4 rows of wires with a and b held constant and only r varies between blocks.  As the lattice constant in the direction of wave propagation has to be less than $0.1 \lambda$ the variations in permittivity and hence QM potential can occur on scales a fraction of a wavelength.  When we consider barriers consisting of differing dielectric slabs such as this example, to represent a spatially changing potential, numerical simulations will have no issue in solving these system, although both QM and EM analytical approaches will struggle to produce results.

The material presented here establishes a framework for verification and validation of microscopic emission models in an easily understandable manner and can be readily expanded to more realistic systems.  In future work, we will carry out V\&V on potentials more representative of atomic emission using this formalism.  Of particular significance, is the use of a macroscopic EM test apparatus to anchor atomistic quantum mechanical emission models.  This work will ultimately enable a solidly vetted atomistic emission model that can be used as part of a multiscale emission model with a high degree of confidence, enabling greater accuracy in the prediction of emitter behavior and the design of optimized emission sources from novel materials.

\section{Dedication}

We would like to dedicate this work to Hertha Ayrton (1854-1923), British engineer, mathematician, physicist, inventor, and suffragette.   This dedication is appropriate given the interdisciplinary nature of her research, which is at the heart of this paper, as well as her tireless efforts throughout her career pushing for the rights of women in society and in science, and leading by example.

\section{Data Availability Statement}

All material necessary to reproduce the results in this paper can be found within the body of the text.

\acknowledgments
The authors would like to thank Dr's Kevin Jensen, Don Shiffler and Prof. Joel Lebowitz for insightful discussions. JR gratefully acknowledges the support provided by the Air Force Office of Scientific Research (AFOSR) through the lab
task 21RDCOR008.  RS gratefully acknowledges support from the AFRL Directed
Energy Chief Scientist Office and the EOARD, grant
FA8655-20-1-7002.

\emph{The views expressed are those of the authors and do not reflect the official guidance or position of the United States Government, the Department of Defense, or of the United States Air Force.}

\emph{Statement from DoD:  The appearance of external hyperlinks does not constitute endorsement by the United States Department of Defense (DoD) of the linked websites, or the information, products, or services contained therein.  The DoD does not exercise any editorial, security, or other control over the information you may find at these locations.}


\appendix

\section{The $\delta$-Function Potential Barrier}
{As with the rectangular barrier, we consider a particle propagating in the positive x-direction becoming incident on a $\delta$-function barrier positioned at $x=0$.  To the right of the barrier is free space.  For this simple illustration, we assume that $V_0 L$ is normalized to unity.}

{Schr\"{o}dinger's equation for this system is:
\begin{equation} \label{eq:SEdelta}
-\frac{\hbar^2}{2m}\frac{d^\psi(x)}{dx^2} + V(x)\psi(x) = E\psi(x)
\end{equation}
where
\begin{equation} \label{deltapotential}
V(x) = \lambda \delta(x) = \left \{ \begin{array}{cc}
    \infty & \mbox{if $x=0$}  \\
    0 & \mbox{if $x\neq 0$}
    \end{array}
    \right.
\end{equation}
In this expression, $\lambda = V_0 L$, which does not change the expression numerically, but ensures that the units work out correctly.}

{The solution of Schr\"{o}dinger's equation on both the left and right sides of the $\delta$-function gives simple plane waves as the particle wavefunction:
\begin{equation} \label{eq:psideltaL}
\psi_L(x) = Ae^{\imath k x} + A'e^{-\imath k x}
\end{equation}
\begin{equation} \label{eq:psideltaR}
\psi_R(x) = Be^{\imath k x} + B'e^{-\imath k x}
\end{equation}
The value of the coefficients can be found using the boundary conditions.  The first boundary condition is obtained by setting $\psi_L(x) = \psi_R(x)$ at $x=0$ to get:
\begin{equation} \label{eq:bound1}
A + A' - B - B' = 0
\end{equation}
The second boundary condition is a little more tricky.  Since there exists a point singularity at $x=0$, the derivative of the wavefunction is undefined here, and the other boundary condition cannot be found by simply setting the first derivatives of the wavefunctions on the left and right equal to one another.}  {To get the other boundary condition, we must go back to the definition of Schr\:{o}dinger's equation for this system and integrate it between $-\epsilon$ and $+\epsilon$:
\begin{equation} \label{intergrateSE}
-\frac{\hbar^2}{2m} \int_{-\epsilon}^{+\epsilon} \psi^{''}(x)dx + \int_{-\epsilon}^{+\epsilon} \lambda \delta(x) \psi(x) dx = E \int_{-\epsilon}^{+\epsilon} \psi(x) dx
\end{equation}
The first term on the left gives the first derivative of the wavefunction at $+\epsilon$ minus the first derivative of the wavefunction at $-\epsilon$:
\begin{equation}
    -\frac{\hbar^2}{2m}(B\imath k e^{\imath k\epsilon} - B'\imath k e^{-\imath k\epsilon} - A\imath k e^{-\imath k\epsilon} + A'\imath k e^{\imath k\epsilon})
\end{equation}
Taking the limit of this as $\epsilon \rightarrow 0$,
\begin{equation}
    -\frac{\imath \hbar^2 k}{2m}(B - B' - A + A')
\end{equation}
The second term on the left simply integrates to $\lambda \psi(0)$ since the integral of the $\delta$-function in these limits is unity.}  {Since the left and right side wavefunctions are equal at $x=0$, we get:
\begin{equation}
    \psi_L(0) = \psi_R(0) = B + B'
\end{equation}
Finally, the term on the right hand side, $E(\psi_R(+\epsilon)-\psi_L(-\epsilon)$, will vanish as $\epsilon \rightarrow 0$.  The second boundary condition is found to be:
\begin{equation} \label{eq:bound2}
-\frac{\imath \hbar^2 k}{2m}(B - B' - A + A') + \lambda(B + B') = 0
\end{equation}
We multiply both sides of the equation by $\frac{1}{\lambda}$ and define a term $\gamma = \frac{k}{\lambda}$.  Recalling that the particle propagates from $-\infty$ and there is only free space to the right of the barrier and therefore nothing travelling to the left, we can rewrite the coefficients as:
\begin{eqnarray}
A = 1 \,\,\,\,\,\, & A' = r \\ \nonumber
B = t \,\,\,\,\,\, & B' = 0
\end{eqnarray}
We now have 2 equations and 2 unknowns, and the boundary conditions can be rewritten as:
\begin{equation}
    1 + r = t
\end{equation}
\begin{equation}
    \frac{\imath \hbar^2 \gamma}{2m}(t+r-1) = t
\end{equation}
Solving for the complex reflection and transmission coefficients, we obtain:
\begin{equation} \label{eq:deltar}
r = \frac{1}{\frac{\imath \hbar^2 \gamma}{m}-1}
\end{equation}
\begin{equation} \label{eq:deltat}
t = \frac{1}{1-\frac{m}{\imath \hbar^2 \gamma}}
\end{equation}
}


\begin{thebibliography}{51}%
\makeatletter
\providecommand \@ifxundefined [1]{%
 \@ifx{#1\undefined}
}%
\providecommand \@ifnum [1]{%
 \ifnum #1\expandafter \@firstoftwo
 \else \expandafter \@secondoftwo
 \fi
}%
\providecommand \@ifx [1]{%
 \ifx #1\expandafter \@firstoftwo
 \else \expandafter \@secondoftwo
 \fi
}%
\providecommand \natexlab [1]{#1}%
\providecommand \enquote  [1]{``#1''}%
\providecommand \bibnamefont  [1]{#1}%
\providecommand \bibfnamefont [1]{#1}%
\providecommand \citenamefont [1]{#1}%
\providecommand \href@noop [0]{\@secondoftwo}%
\providecommand \href [0]{\begingroup \@sanitize@url \@href}%
\providecommand \@href[1]{\@@startlink{#1}\@@href}%
\providecommand \@@href[1]{\endgroup#1\@@endlink}%
\providecommand \@sanitize@url [0]{\catcode `\\12\catcode `\$12\catcode
  `\&12\catcode `\#12\catcode `\^12\catcode `\_12\catcode `\%12\relax}%
\providecommand \@@startlink[1]{}%
\providecommand \@@endlink[0]{}%
\providecommand \url  [0]{\begingroup\@sanitize@url \@url }%
\providecommand \@url [1]{\endgroup\@href {#1}{\urlprefix }}%
\providecommand \urlprefix  [0]{URL }%
\providecommand \Eprint [0]{\href }%
\providecommand \doibase [0]{https://doi.org/}%
\providecommand \selectlanguage [0]{\@gobble}%
\providecommand \bibinfo  [0]{\@secondoftwo}%
\providecommand \bibfield  [0]{\@secondoftwo}%
\providecommand \translation [1]{[#1]}%
\providecommand \BibitemOpen [0]{}%
\providecommand \bibitemStop [0]{}%
\providecommand \bibitemNoStop [0]{.\EOS\space}%
\providecommand \EOS [0]{\spacefactor3000\relax}%
\providecommand \BibitemShut  [1]{\csname bibitem#1\endcsname}%
\let\auto@bib@innerbib\@empty
\bibitem [{\citenamefont {Harris}\ \emph {et~al.}(2017)\citenamefont {Harris},
  \citenamefont {Jensen}, \citenamefont {Petillo}, \citenamefont {Maestas},
  \citenamefont {Tang},\ and\ \citenamefont {Shiffler}}]{Harris2017}%
  \BibitemOpen
  \bibfield  {author} {\bibinfo {author} {\bibfnamefont {J.~R.}\ \bibnamefont
  {Harris}}, \bibinfo {author} {\bibfnamefont {K.~L.}\ \bibnamefont {Jensen}},
  \bibinfo {author} {\bibfnamefont {J.~J.}\ \bibnamefont {Petillo}}, \bibinfo
  {author} {\bibfnamefont {S.}~\bibnamefont {Maestas}}, \bibinfo {author}
  {\bibfnamefont {W.}~\bibnamefont {Tang}},\ and\ \bibinfo {author}
  {\bibfnamefont {D.~A.}\ \bibnamefont {Shiffler}},\ }\href
  {https://doi.org/10.1063/1.4983680} {\bibfield  {journal} {\bibinfo
  {journal} {J. Appl. Phys.}\ }\textbf {\bibinfo {volume} {121}},\ \bibinfo
  {pages} {203303} (\bibinfo {year} {2017})}\BibitemShut {NoStop}%
\bibitem [{\citenamefont {Harris}(2018)}]{Harris2018}%
  \BibitemOpen
  \bibfield  {author} {\bibinfo {author} {\bibfnamefont {J.~R.}\ \bibnamefont
  {Harris}},\ }\href {https://doi.org/10.1109/TPS.2017.2759248} {\bibfield
  {journal} {\bibinfo  {journal} {IEEE Trans. Plasma Sci.}\ }\textbf {\bibinfo
  {volume} {46}},\ \bibinfo {pages} {1872} (\bibinfo {year}
  {2018})}\BibitemShut {NoStop}%
\bibitem [{\citenamefont {Schottky}(1923)}]{Schottky1923}%
  \BibitemOpen
  \bibfield  {author} {\bibinfo {author} {\bibfnamefont {W.}~\bibnamefont
  {Schottky}},\ }\href {https://doi.org/10.1007/BF01340034} {\bibfield
  {journal} {\bibinfo  {journal} {Z. Phys.}\ }\textbf {\bibinfo {volume}
  {14}},\ \bibinfo {pages} {63} (\bibinfo {year} {1923})}\BibitemShut {NoStop}%
\bibitem [{\citenamefont {Stern}\ \emph {et~al.}(1929)\citenamefont {Stern},
  \citenamefont {Gossling},\ and\ \citenamefont {Fowler}}]{Stern1929}%
  \BibitemOpen
  \bibfield  {author} {\bibinfo {author} {\bibfnamefont {T.}~\bibnamefont
  {Stern}}, \bibinfo {author} {\bibfnamefont {B.}~\bibnamefont {Gossling}},\
  and\ \bibinfo {author} {\bibfnamefont {R.}~\bibnamefont {Fowler}},\ }\href
  {https://doi.org/10.1098/rspa.1929.0147} {\bibfield  {journal} {\bibinfo
  {journal} {Proc. R. Soc. A}\ }\textbf {\bibinfo {volume} {124}},\ \bibinfo
  {pages} {699} (\bibinfo {year} {1929})}\BibitemShut {NoStop}%
\bibitem [{\citenamefont {Miller}\ \emph {et~al.}(2009)\citenamefont {Miller},
  \citenamefont {Lau},\ and\ \citenamefont {Booske}}]{Miller2009}%
  \BibitemOpen
  \bibfield  {author} {\bibinfo {author} {\bibfnamefont {R.}~\bibnamefont
  {Miller}}, \bibinfo {author} {\bibfnamefont {Y.}~\bibnamefont {Lau}},\ and\
  \bibinfo {author} {\bibfnamefont {J.}~\bibnamefont {Booske}},\ }\href
  {https://doi.org/10.1063/1.3253760} {\bibfield  {journal} {\bibinfo
  {journal} {J. Appl. Phys.}\ }\textbf {\bibinfo {volume} {106}},\ \bibinfo
  {pages} {104903} (\bibinfo {year} {2009})}\BibitemShut {NoStop}%
\bibitem [{\citenamefont {Harris}\ \emph {et~al.}(2015)\citenamefont {Harris},
  \citenamefont {Jensen},\ and\ \citenamefont {Shiffler}}]{Harris2015}%
  \BibitemOpen
  \bibfield  {author} {\bibinfo {author} {\bibfnamefont {J.~R.}\ \bibnamefont
  {Harris}}, \bibinfo {author} {\bibfnamefont {K.~L.}\ \bibnamefont {Jensen}},\
  and\ \bibinfo {author} {\bibfnamefont {D.~A.}\ \bibnamefont {Shiffler}},\
  }\href {https://doi.org/10.1088/0022-3727/48/38/385203} {\bibfield  {journal}
  {\bibinfo  {journal} {J. Phys. D Appl. Phys.}\ }\textbf {\bibinfo {volume}
  {48}},\ \bibinfo {pages} {385203} (\bibinfo {year} {2015})}\BibitemShut
  {NoStop}%
\bibitem [{\citenamefont {Jensen}\ \emph {et~al.}(2016)\citenamefont {Jensen},
  \citenamefont {Shiffler}, \citenamefont {Harris},\ and\ \citenamefont
  {Petillo}}]{Jensen2016}%
  \BibitemOpen
  \bibfield  {author} {\bibinfo {author} {\bibfnamefont {K.~L.}\ \bibnamefont
  {Jensen}}, \bibinfo {author} {\bibfnamefont {D.~A.}\ \bibnamefont
  {Shiffler}}, \bibinfo {author} {\bibfnamefont {J.~R.}\ \bibnamefont
  {Harris}},\ and\ \bibinfo {author} {\bibfnamefont {J.~J.}\ \bibnamefont
  {Petillo}},\ }\href {https://doi.org/10.1063/1.4953813} {\bibfield  {journal}
  {\bibinfo  {journal} {AIP Adv.}\ }\textbf {\bibinfo {volume} {6}},\ \bibinfo
  {pages} {065005} (\bibinfo {year} {2016})}\BibitemShut {NoStop}%
\bibitem [{\citenamefont {Kyritsakis}\ \emph {et~al.}(2010)\citenamefont
  {Kyritsakis}, \citenamefont {Kokkorakis}, \citenamefont {Xanthakis},
  \citenamefont {Kirk},\ and\ \citenamefont {Pescia}}]{Kyritsakis2010}%
  \BibitemOpen
  \bibfield  {author} {\bibinfo {author} {\bibfnamefont {A.}~\bibnamefont
  {Kyritsakis}}, \bibinfo {author} {\bibfnamefont {G.}~\bibnamefont
  {Kokkorakis}}, \bibinfo {author} {\bibfnamefont {J.}~\bibnamefont
  {Xanthakis}}, \bibinfo {author} {\bibfnamefont {T.}~\bibnamefont {Kirk}},\
  and\ \bibinfo {author} {\bibfnamefont {D.}~\bibnamefont {Pescia}},\ }\href
  {https://doi.org/10.1063/1.3462934} {\bibfield  {journal} {\bibinfo
  {journal} {Appl. Phys. Lett.}\ }\textbf {\bibinfo {volume} {97}},\ \bibinfo
  {pages} {023104 / 1} (\bibinfo {year} {2010})}\BibitemShut {NoStop}%
\bibitem [{\citenamefont {Kyritsakis}\ and\ \citenamefont
  {Xanthakis}(2015)}]{Kyritsakis2015}%
  \BibitemOpen
  \bibfield  {author} {\bibinfo {author} {\bibfnamefont {A.}~\bibnamefont
  {Kyritsakis}}\ and\ \bibinfo {author} {\bibfnamefont {J.~P.}\ \bibnamefont
  {Xanthakis}},\ }\href {https://doi.org/10.1098/rspa.2014.0811} {\bibfield
  {journal} {\bibinfo  {journal} {Proc. R. Soc. A}\ }\textbf {\bibinfo {volume}
  {471}},\ \bibinfo {pages} {20140811} (\bibinfo {year} {2015})}\BibitemShut
  {NoStop}%
\bibitem [{\citenamefont {Jensen}\ \emph {et~al.}(2017)\citenamefont {Jensen},
  \citenamefont {Shiffler}, \citenamefont {Peckerar}, \citenamefont {Harris},\
  and\ \citenamefont {Petillo}}]{Jensen2017}%
  \BibitemOpen
  \bibfield  {author} {\bibinfo {author} {\bibfnamefont {K.~L.}\ \bibnamefont
  {Jensen}}, \bibinfo {author} {\bibfnamefont {D.~A.}\ \bibnamefont
  {Shiffler}}, \bibinfo {author} {\bibfnamefont {M.}~\bibnamefont {Peckerar}},
  \bibinfo {author} {\bibfnamefont {J.~R.}\ \bibnamefont {Harris}},\ and\
  \bibinfo {author} {\bibfnamefont {J.~J.}\ \bibnamefont {Petillo}},\ }\href
  {https://doi.org/10.1063/1.4997457} {\bibfield  {journal} {\bibinfo
  {journal} {J. Appl. Phys.}\ }\textbf {\bibinfo {volume} {122}},\ \bibinfo
  {pages} {064501} (\bibinfo {year} {2017})}\BibitemShut {NoStop}%
\bibitem [{\citenamefont {Tang}\ \emph {et~al.}(2014)\citenamefont {Tang},
  \citenamefont {Shiffler}, \citenamefont {Golby}, \citenamefont {LaCour},\
  and\ \citenamefont {Knowles}}]{Tang2014}%
  \BibitemOpen
  \bibfield  {author} {\bibinfo {author} {\bibfnamefont {W.}~\bibnamefont
  {Tang}}, \bibinfo {author} {\bibfnamefont {D.}~\bibnamefont {Shiffler}},
  \bibinfo {author} {\bibfnamefont {K.}~\bibnamefont {Golby}}, \bibinfo
  {author} {\bibfnamefont {M.}~\bibnamefont {LaCour}},\ and\ \bibinfo {author}
  {\bibfnamefont {T.}~\bibnamefont {Knowles}},\ }\href
  {https://doi.org/10.1116/1.4891928} {\bibfield  {journal} {\bibinfo
  {journal} {J. Vac. Sci. Technol. B}\ }\textbf {\bibinfo {volume} {32}},\
  \bibinfo {pages} {052202} (\bibinfo {year} {2014})}\BibitemShut {NoStop}%
\bibitem [{\citenamefont {Binh}\ \emph {et~al.}(1992)\citenamefont {Binh},
  \citenamefont {Purcell}, \citenamefont {Garcia},\ and\ \citenamefont
  {Doglioni}}]{Binh1992}%
  \BibitemOpen
  \bibfield  {author} {\bibinfo {author} {\bibfnamefont {V.~T.}\ \bibnamefont
  {Binh}}, \bibinfo {author} {\bibfnamefont {S.~T.}\ \bibnamefont {Purcell}},
  \bibinfo {author} {\bibfnamefont {N.}~\bibnamefont {Garcia}},\ and\ \bibinfo
  {author} {\bibfnamefont {J.}~\bibnamefont {Doglioni}},\ }\href
  {https://doi.org/10.1103/PhysRevLett.69.2527} {\bibfield  {journal} {\bibinfo
   {journal} {Phys. Rev. Lett.}\ }\textbf {\bibinfo {volume} {69}},\ \bibinfo
  {pages} {2527} (\bibinfo {year} {1992})}\BibitemShut {NoStop}%
\bibitem [{\citenamefont {Shaw}\ and\ \citenamefont {Itoh}(2001)}]{Zhu2001}%
  \BibitemOpen
  \bibfield  {author} {\bibinfo {author} {\bibfnamefont {J.}~\bibnamefont
  {Shaw}}\ and\ \bibinfo {author} {\bibfnamefont {J.}~\bibnamefont {Itoh}},\
  }in\ \href@noop {} {\emph {\bibinfo {booktitle} {Vacuum Microelectronics}}},\
  \bibinfo {editor} {edited by\ \bibinfo {editor} {\bibfnamefont
  {W.}~\bibnamefont {Zhu}}}\ (\bibinfo  {publisher} {Wiley},\ \bibinfo
  {address} {New York},\ \bibinfo {year} {2001})\ p.\ \bibinfo {pages}
  {187}\BibitemShut {NoStop}%
\bibitem [{\citenamefont {Gallmann}\ \emph {et~al.}(2017)\citenamefont
  {Gallmann}, \citenamefont {Jordan}, \citenamefont {W{\"o}rner}, \citenamefont
  {Castiglioni}, \citenamefont {Hengsberger}, \citenamefont {Osterwalder},
  \citenamefont {Arrell}, \citenamefont {Chergui}, \citenamefont {Liberatore},
  \citenamefont {Rothlisberger},\ and\ \citenamefont {Keller}}]{Gallmann2017}%
  \BibitemOpen
  \bibfield  {author} {\bibinfo {author} {\bibfnamefont {L.}~\bibnamefont
  {Gallmann}}, \bibinfo {author} {\bibfnamefont {I.}~\bibnamefont {Jordan}},
  \bibinfo {author} {\bibfnamefont {H.~J.}\ \bibnamefont {W{\"o}rner}},
  \bibinfo {author} {\bibfnamefont {L.}~\bibnamefont {Castiglioni}}, \bibinfo
  {author} {\bibfnamefont {M.}~\bibnamefont {Hengsberger}}, \bibinfo {author}
  {\bibfnamefont {J.}~\bibnamefont {Osterwalder}}, \bibinfo {author}
  {\bibfnamefont {C.~A.}\ \bibnamefont {Arrell}}, \bibinfo {author}
  {\bibfnamefont {M.}~\bibnamefont {Chergui}}, \bibinfo {author} {\bibfnamefont
  {E.}~\bibnamefont {Liberatore}}, \bibinfo {author} {\bibfnamefont
  {U.}~\bibnamefont {Rothlisberger}},\ and\ \bibinfo {author} {\bibfnamefont
  {U.}~\bibnamefont {Keller}},\ }\href {https://doi.org/10.1063/1.4997175}
  {\bibfield  {journal} {\bibinfo  {journal} {Structural Dynamics}\ }\textbf
  {\bibinfo {volume} {4}},\ \bibinfo {pages} {061502} (\bibinfo {year}
  {2017})}\BibitemShut {NoStop}%
\bibitem [{\citenamefont {Go}\ and\ \citenamefont {Pohlman}(2010)}]{Go2010}%
  \BibitemOpen
  \bibfield  {author} {\bibinfo {author} {\bibfnamefont {D.~B.}\ \bibnamefont
  {Go}}\ and\ \bibinfo {author} {\bibfnamefont {D.~A.}\ \bibnamefont
  {Pohlman}},\ }\href {https://doi.org/10.1063/1.3380855} {\bibfield  {journal}
  {\bibinfo  {journal} {J. Appl. Phys.}\ }\textbf {\bibinfo {volume} {107}},\
  \bibinfo {pages} {103303} (\bibinfo {year} {2010})}\BibitemShut {NoStop}%
\bibitem [{\citenamefont {Edelen}\ \emph {et~al.}(2015)\citenamefont {Edelen},
  \citenamefont {Biedron}, \citenamefont {Harris}, \citenamefont {Milton},\
  and\ \citenamefont {Lewellen}}]{Edelen2015}%
  \BibitemOpen
  \bibfield  {author} {\bibinfo {author} {\bibfnamefont {J.~P.}\ \bibnamefont
  {Edelen}}, \bibinfo {author} {\bibfnamefont {S.~G.}\ \bibnamefont {Biedron}},
  \bibinfo {author} {\bibfnamefont {J.~R.}\ \bibnamefont {Harris}}, \bibinfo
  {author} {\bibfnamefont {S.~V.}\ \bibnamefont {Milton}},\ and\ \bibinfo
  {author} {\bibfnamefont {J.~W.}\ \bibnamefont {Lewellen}},\ }\href
  {https://doi.org/10.1103/PhysRevSTAB.18.043402} {\bibfield  {journal}
  {\bibinfo  {journal} {Phys. Rev. ST Accel. Beams}\ }\textbf {\bibinfo
  {volume} {18}},\ \bibinfo {pages} {043402} (\bibinfo {year}
  {2015})}\BibitemShut {NoStop}%
\bibitem [{\citenamefont {Dynako}\ \emph {et~al.}(2018)\citenamefont {Dynako},
  \citenamefont {Loveless},\ and\ \citenamefont {Garner}}]{Dynako2018}%
  \BibitemOpen
  \bibfield  {author} {\bibinfo {author} {\bibfnamefont {S.~D.}\ \bibnamefont
  {Dynako}}, \bibinfo {author} {\bibfnamefont {A.~M.}\ \bibnamefont
  {Loveless}},\ and\ \bibinfo {author} {\bibfnamefont {A.~L.}\ \bibnamefont
  {Garner}},\ }\href@noop {} {\bibfield  {journal} {\bibinfo  {journal} {Phys.
  Plasmas}\ }\textbf {\bibinfo {volume} {25}} (\bibinfo {year}
  {2018})}\BibitemShut {NoStop}%
\bibitem [{\citenamefont {Jensen}\ \emph
  {et~al.}(2019{\natexlab{a}})\citenamefont {Jensen}, \citenamefont {Shiffler},
  \citenamefont {Lebowitz}, \citenamefont {Cahay},\ and\ \citenamefont
  {Petillo}}]{Jensen2019}%
  \BibitemOpen
  \bibfield  {author} {\bibinfo {author} {\bibfnamefont {K.~L.}\ \bibnamefont
  {Jensen}}, \bibinfo {author} {\bibfnamefont {D.~A.}\ \bibnamefont
  {Shiffler}}, \bibinfo {author} {\bibfnamefont {J.~L.}\ \bibnamefont
  {Lebowitz}}, \bibinfo {author} {\bibfnamefont {M.}~\bibnamefont {Cahay}},\
  and\ \bibinfo {author} {\bibfnamefont {J.~J.}\ \bibnamefont {Petillo}},\
  }\href {https://doi.org/10.1063/1.5086434} {\bibfield  {journal} {\bibinfo
  {journal} {J. Appl. Phys.}\ }\textbf {\bibinfo {volume} {125}},\ \bibinfo
  {pages} {114303} (\bibinfo {year} {2019}{\natexlab{a}})}\BibitemShut
  {NoStop}%
\bibitem [{\citenamefont {Jensen}\ \emph
 (2018)\citenamefont {Jensen}}]{Jensen2018}%
  \BibitemOpen
  \bibfield  {author} {\bibinfo {author} {\bibfnamefont {K.~L.}\ \bibnamefont
  {Jensen}},\
  }\href {https://doi.org/10.1109/TPS.2017.2782485} {\bibfield  {journal} {\bibinfo
  {journal} {IEEE Trans. Plas. Sci.}\ }\textbf {\bibinfo {volume} {46}},\ \bibinfo
{pages} {1881-1899} (\bibinfo {year} {2018}{\natexlab{a}})}\BibitemShut
  {NoStop}%
\bibitem [{\citenamefont {Jensen}\ \emph
 {et~al.}(2019{\natexlab{a}})\citenamefont {Jensen}, \citenamefont {Shabaev}, \citenamefont {Lambrakos}, \citenamefont {Finkenstadt}, \citenamefont {Moody}, \citenamefont{Neukirch}, \citenamefont{Tretiak}, \citenamefont{Shiffler},\ and\ \citenamefont{Petillo}}]{Jensen2020}%
  \BibitemOpen
  \bibfield  {author} {\bibinfo {author} {\bibfnamefont {K.~L.}\ \bibnamefont
  {Jensen}},\bibinfo {author} {\bibnamefont{A.}\ \bibnamefont{Shabaev}},\bibinfo {author} {\bibfnamefont {S.~G.}\ \bibnamefont
  {Lambrakos}},\bibinfo {author} {\bibfnamefont {D.}\ \bibnamefont
  {Finkenstadt}},\bibinfo {author} {\bibfnamefont {N.~A.}\ \bibnamefont
  {Moody}},\bibinfo {author} {\bibfnamefont {A.~J.}\ \bibnamefont
  {Neukirch}},\bibinfo {author} {\bibfnamefont {S.}\ \bibnamefont
  {Tretiak}},\bibinfo {author} {\bibfnamefont {D.~A.}\ \bibnamefont
  {Shiffler}},\bibinfo {author} {\bibfnamefont {J.~J.}\ \bibnamefont
  {Petillo}}, 
  }\href {http://dx.doi.org/10.1063/5.0009759} {\bibfield  {journal} {\bibinfo
  {journal} {J. Appl. Phys.}\ }\textbf {\bibinfo {volume} {127}},\ \bibinfo
  {pages} {235301} (\bibinfo {year} {2020}{\natexlab{a}})}\BibitemShut
  {NoStop}%
\bibitem [{\citenamefont {Ando}\ \emph
 {et~al.}(1998{\natexlab{a}})\citenamefont {Ando}, \citenamefont {Itoh}}]{Ando1998}%
  \BibitemOpen
  \bibfield  {author} {\bibinfo {author} {\bibfnamefont {Y.}\ \bibnamefont
  {Ando}},\bibinfo {author} {\bibnamefont{T.}\ \bibnamefont{Itoh}}, 
  }\href {https://doi.org/10.1063/1.338082} {\bibfield  {journal} {\bibinfo
  {journal} {J. Appl. Phys.}\ }\textbf {\bibinfo {volume} {61}},\ \bibinfo
  {pages} {1497} (\bibinfo {year} {1998}{\natexlab{a}})}\BibitemShut
  {NoStop}%
\bibitem [{\citenamefont {Mayer}\ \emph
 (2010{\natexlab{a}})\citenamefont {Mayer}}]{Mayer2010}%
  \BibitemOpen
  \bibfield  {author} {\bibinfo {author} {\bibfnamefont {A.}\ \bibnamefont
  {Mayer}},
  }\href {https://doi.org/10.1088/0953-8984/22/17/175007} {\bibfield  {journal} {\bibinfo
  {journal} {J. Phys.: Condens. Matter}\ }\textbf {\bibinfo {volume} {22}},\ \bibinfo
  {pages} {175007} (\bibinfo {year} {2010}{\natexlab{a}})}\BibitemShut
  {NoStop}%
\bibitem [{\citenamefont {Ludwick}\ \emph
 {et~al.}(1998{\natexlab{a}})\citenamefont {Ludwick}, \citenamefont {Cahay}, \citenamefont {Hernandez}}]{Ludwick2021}%
  \BibitemOpen
  \bibfield  {author} {\bibinfo {author} {\bibfnamefont {J.}\ \bibnamefont
  {Ludwick}},\bibinfo {author} {\bibnamefont{M.}\ \bibnamefont{Cahay}},\bibinfo {author} {\bibnamefont{N.}\ \bibnamefont{Hernandez}}, 
  }\href {https://doi.org/10.1063/5.0065612} {\bibfield  {journal} {\bibinfo
  {journal} {J. Appl. Phys.}\ }\textbf {\bibinfo {volume} {130}},\ \bibinfo
  {pages} {144302} (\bibinfo {year} {2021}{\natexlab{a}})}\BibitemShut
  {NoStop}%
\bibitem [{\citenamefont {Jensen}\ \emph
  {et~al.}(2019{\natexlab{b}})\citenamefont {Jensen}, \citenamefont {Shiffler},
  \citenamefont {Riga}, \citenamefont {Harris}, \citenamefont {Lebowitz},
  \citenamefont {Cahay},\ and\ \citenamefont {Petillo}}]{Jensen2019a}%
  \BibitemOpen
  \bibfield  {author} {\bibinfo {author} {\bibfnamefont {K.~L.}\ \bibnamefont
  {Jensen}}, \bibinfo {author} {\bibfnamefont {D.~A.}\ \bibnamefont
  {Shiffler}}, \bibinfo {author} {\bibfnamefont {J.~M.}\ \bibnamefont {Riga}},
  \bibinfo {author} {\bibfnamefont {J.~R.}\ \bibnamefont {Harris}}, \bibinfo
  {author} {\bibfnamefont {J.~L.}\ \bibnamefont {Lebowitz}}, \bibinfo {author}
  {\bibfnamefont {M.}~\bibnamefont {Cahay}},\ and\ \bibinfo {author}
  {\bibfnamefont {J.~J.}\ \bibnamefont {Petillo}},\ }\href
  {https://doi.org/10.1063/1.5110406} {\bibfield  {journal} {\bibinfo
  {journal} {J. Appl. Phys.}\ }\textbf {\bibinfo {volume} {126}},\ \bibinfo
  {pages} {144301} (\bibinfo {year} {2019}{\natexlab{b}})}\BibitemShut
  {NoStop}%
\bibitem [{\citenamefont {Jensen}\ \emph
  {et~al.}(2021{\natexlab{a}})\citenamefont {Jensen}, \citenamefont {Lebowitz},
  \citenamefont {Riga}, \citenamefont {Shiffler},\ and\ \citenamefont
  {Seviour}}]{Jensen2021}%
  \BibitemOpen
  \bibfield  {author} {\bibinfo {author} {\bibfnamefont {K.~L.}\ \bibnamefont
  {Jensen}}, \bibinfo {author} {\bibfnamefont {J.~L.}\ \bibnamefont
  {Lebowitz}}, \bibinfo {author} {\bibfnamefont {J.~M.}\ \bibnamefont {Riga}},
  \bibinfo {author} {\bibfnamefont {D.~A.}\ \bibnamefont {Shiffler}},\ and\
  \bibinfo {author} {\bibfnamefont {R.}~\bibnamefont {Seviour}},\ }\href
  {https://doi.org/10.1103/PhysRevB.103.155427} {\bibfield  {journal} {\bibinfo
   {journal} {Phys. Rev. B}\ }\textbf {\bibinfo {volume} {103}},\ \bibinfo
  {pages} {155427} (\bibinfo {year} {2021}{\natexlab{a}})}\BibitemShut
  {NoStop}%
\bibitem [{\citenamefont {Jensen}\ \emph
  {et~al.}(2021{\natexlab{b}})\citenamefont {Jensen}, \citenamefont {Riga},
  \citenamefont {Shiffler}, \citenamefont {Lebowitz},\ and\ \citenamefont
  {Seviour}}]{Jensen2021a}%
  \BibitemOpen
  \bibfield  {author} {\bibinfo {author} {\bibfnamefont {K.~L.}\ \bibnamefont
  {Jensen}}, \bibinfo {author} {\bibfnamefont {J.~M.}\ \bibnamefont {Riga}},
  \bibinfo {author} {\bibfnamefont {D.~A.}\ \bibnamefont {Shiffler}}, \bibinfo
  {author} {\bibfnamefont {J.~L.}\ \bibnamefont {Lebowitz}},\ and\ \bibinfo
  {author} {\bibfnamefont {R.}~\bibnamefont {Seviour}},\ }\href
  {https://doi.org/10.1103/PhysRevA.104.062203} {\bibfield  {journal} {\bibinfo
   {journal} {Phys. Rev. A}\ }\textbf {\bibinfo {volume} {104}},\ \bibinfo
  {pages} {062203} (\bibinfo {year} {2021}{\natexlab{b}})}\BibitemShut
  {NoStop}%
\bibitem [{\citenamefont {Jensen}\ \emph {et~al.}(2022)\citenamefont {Jensen},
  \citenamefont {Riga}, \citenamefont {Lebowitz}, \citenamefont {Seviour},\
  and\ \citenamefont {Shiffler}}]{Jensen2022}%
  \BibitemOpen
  \bibfield  {author} {\bibinfo {author} {\bibfnamefont {K.~L.}\ \bibnamefont
  {Jensen}}, \bibinfo {author} {\bibfnamefont {J.}~\bibnamefont {Riga}},
  \bibinfo {author} {\bibfnamefont {J.~L.}\ \bibnamefont {Lebowitz}}, \bibinfo
  {author} {\bibfnamefont {R.}~\bibnamefont {Seviour}},\ and\ \bibinfo {author}
  {\bibfnamefont {D.~A.}\ \bibnamefont {Shiffler}},\ }\href
  {https://doi.org/10.1063/5.0096568} {\bibfield  {journal} {\bibinfo
  {journal} {Journal of Applied Physics}\ }\textbf {\bibinfo {volume} {132}},\
  \bibinfo {pages} {124303} (\bibinfo {year} {2022})}\BibitemShut {NoStop}%
\bibitem [{\citenamefont {Landauer}(1990)}]{Rolf}%
  \BibitemOpen
  \bibfield  {author} {\bibinfo {author} {\bibfnamefont {R.}~\bibnamefont
  {Landauer}},\ }\href@noop {} {\emph {\bibinfo {title} {Analogies in Optics
  and Micro Electronics}}}\ (\bibinfo  {publisher} {Kluwer Academic},\ \bibinfo
  {address} {Amsterdam, Netherlands},\ \bibinfo {year} {1990})\ p.\ \bibinfo
  {pages} {243}\BibitemShut {NoStop}%
\bibitem [{\citenamefont {Shockley}(1950)}]{Shok}%
  \BibitemOpen
  \bibfield  {author} {\bibinfo {author} {\bibfnamefont {W.}~\bibnamefont
  {Shockley}},\ }\href@noop {} {\emph {\bibinfo {title} {Electrons and Holes in
  Semiconductors}}}\ (\bibinfo  {publisher} {Van Nostrand},\ \bibinfo {address}
  {New York, USA},\ \bibinfo {year} {1950})\ p.\ \bibinfo {pages} {Chap
  14}\BibitemShut {NoStop}%
\bibitem [{\citenamefont {Dragoman}\ and\ \citenamefont
  {Dragoman}(2013)}]{Drago}%
  \BibitemOpen
  \bibfield  {author} {\bibinfo {author} {\bibfnamefont {D.}~\bibnamefont
  {Dragoman}}\ and\ \bibinfo {author} {\bibfnamefont {M.}~\bibnamefont
  {Dragoman}},\ }\href@noop {} {\emph {\bibinfo {title} {Quantum-Classical
  Analogies (The Frontiers Collection}}}\ (\bibinfo  {publisher} {Springer},\
  \bibinfo {address} {Berlin},\ \bibinfo {year} {2013})\ p.\ \bibinfo {pages}
  {354}\BibitemShut {NoStop}%
\bibitem [{\citenamefont {Gersten}(2001)}]{Ger}%
  \BibitemOpen
  \bibfield  {author} {\bibinfo {author} {\bibfnamefont {A.}~\bibnamefont
  {Gersten}},\ }\href {https://doi.org/10.1023/A:1017551920941} {\bibfield
  {journal} {\bibinfo  {journal} {Foundations of Physics}\ }\textbf {\bibinfo
  {volume} {31}},\ \bibinfo {pages} {1211} (\bibinfo {year}
  {2001})}\BibitemShut {NoStop}%
\bibitem [{\citenamefont {Eberly}\ \emph {et~al.}(2013)\citenamefont {Eberly},
  \citenamefont {Mandel},\ and\ \citenamefont {Wolf}}]{Eber}%
  \BibitemOpen
  \bibfield  {author} {\bibinfo {author} {\bibfnamefont {J.}~\bibnamefont
  {Eberly}}, \bibinfo {author} {\bibfnamefont {L.}~\bibnamefont {Mandel}},\
  and\ \bibinfo {author} {\bibfnamefont {E.}~\bibnamefont {Wolf}},\ }\href@noop
  {} {\emph {\bibinfo {title} {Coherence and Quantum Optics VII, Proceedings of
  the Seventh Rochester Conference on Coherence and Quantum Optics}}}\
  (\bibinfo  {publisher} {Plenum Press},\ \bibinfo {address} {New York},\
  \bibinfo {year} {2013})\ p.\ \bibinfo {pages} {313}\BibitemShut {NoStop}%
\bibitem [{\citenamefont {Deutsch}\ and\ \citenamefont {Garrison}(1991)}]{Deu}%
  \BibitemOpen
  \bibfield  {author} {\bibinfo {author} {\bibfnamefont {I.~H.}\ \bibnamefont
  {Deutsch}}\ and\ \bibinfo {author} {\bibfnamefont {J.~C.}\ \bibnamefont
  {Garrison}},\ }\href
  {https://doi.org/https://doi.org/10.1103/PhysRevA.43.2498} {\bibfield
  {journal} {\bibinfo  {journal} {Phys. Rev. A}\ }\textbf {\bibinfo {volume}
  {43}},\ \bibinfo {pages} {2498} (\bibinfo {year} {1991})}\BibitemShut
  {NoStop}%
\bibitem [{\citenamefont {Hanson}(2020)}]{hans}%
  \BibitemOpen
  \bibfield  {author} {\bibinfo {author} {\bibfnamefont {G.}~\bibnamefont
  {Hanson}},\ }\bibfield  {journal} {\bibinfo  {journal} {IEEE Antennas and
  Propagation Magazine}\ }\textbf {\bibinfo {volume} {PP}},\ \href
  {https://doi.org/10.1109/MAP.2020.2990065} {10.1109/MAP.2020.2990065}
  (\bibinfo {year} {2020})\BibitemShut {NoStop}%
\bibitem [{\citenamefont {Lundstrom}\ and\ \citenamefont
  {Datta}(1990)}]{lunds}%
  \BibitemOpen
  \bibfield  {author} {\bibinfo {author} {\bibfnamefont {M.}~\bibnamefont
  {Lundstrom}}\ and\ \bibinfo {author} {\bibfnamefont {S.}~\bibnamefont
  {Datta}},\ }\href@noop {} {\bibfield  {journal} {\bibinfo  {journal} {IEEE
  Circuits Dev.}\ }\textbf {\bibinfo {volume} {6}},\ \bibinfo {pages} {32}
  (\bibinfo {year} {1990})}\BibitemShut {NoStop}%
\bibitem [{\citenamefont {Horsley}(2018)}]{Hor}%
  \BibitemOpen
  \bibfield  {author} {\bibinfo {author} {\bibfnamefont {S.~A.~R.}\
  \bibnamefont {Horsley}},\ }\href {https://doi.org/10.1103/PhysRevA.98.043837}
  {\bibfield  {journal} {\bibinfo  {journal} {Phys. Rev. A}\ }\textbf {\bibinfo
  {volume} {98}},\ \bibinfo {pages} {043837} (\bibinfo {year}
  {2018})}\BibitemShut {NoStop}%
\bibitem [{\citenamefont {Bittner}\ \emph {et~al.}(2010)\citenamefont
  {Bittner}, \citenamefont {Dietz}, \citenamefont {Miski-Oglu}, \citenamefont
  {Oria~Iriarte}, \citenamefont {Richter},\ and\ \citenamefont
  {Sch\"afer}}]{Bit}%
  \BibitemOpen
  \bibfield  {author} {\bibinfo {author} {\bibfnamefont {S.}~\bibnamefont
  {Bittner}}, \bibinfo {author} {\bibfnamefont {B.}~\bibnamefont {Dietz}},
  \bibinfo {author} {\bibfnamefont {M.}~\bibnamefont {Miski-Oglu}}, \bibinfo
  {author} {\bibfnamefont {P.}~\bibnamefont {Oria~Iriarte}}, \bibinfo {author}
  {\bibfnamefont {A.}~\bibnamefont {Richter}},\ and\ \bibinfo {author}
  {\bibfnamefont {F.}~\bibnamefont {Sch\"afer}},\ }\href
  {https://doi.org/10.1103/PhysRevB.82.014301} {\bibfield  {journal} {\bibinfo
  {journal} {Phys. Rev. B}\ }\textbf {\bibinfo {volume} {82}},\ \bibinfo
  {pages} {014301} (\bibinfo {year} {2010})}\BibitemShut {NoStop}%
\bibitem [{\citenamefont {Gaylord}\ \emph {et~al.}(1989)\citenamefont
  {Gaylord}, \citenamefont {Glytsis},\ and\ \citenamefont {Brennan}}]{gay}%
  \BibitemOpen
  \bibfield  {author} {\bibinfo {author} {\bibfnamefont {T.~K.}\ \bibnamefont
  {Gaylord}}, \bibinfo {author} {\bibfnamefont {E.~N.}\ \bibnamefont
  {Glytsis}},\ and\ \bibinfo {author} {\bibfnamefont {K.~F.}\ \bibnamefont
  {Brennan}},\ }\href {https://doi.org/10.1063/1.342775} {\bibfield  {journal}
  {\bibinfo  {journal} {J. Appl. Phys.}\ }\textbf {\bibinfo {volume} {65}},\
  \bibinfo {pages} {2535} (\bibinfo {year} {1989})}\BibitemShut {NoStop}%
\bibitem [{\citenamefont {Glytsis}\ \emph {et~al.}(1989)\citenamefont
  {Glytsis}, \citenamefont {Gaylord},\ and\ \citenamefont {Brennan}}]{glt}%
  \BibitemOpen
  \bibfield  {author} {\bibinfo {author} {\bibfnamefont {E.~N.}\ \bibnamefont
  {Glytsis}}, \bibinfo {author} {\bibfnamefont {T.~K.}\ \bibnamefont
  {Gaylord}},\ and\ \bibinfo {author} {\bibfnamefont {K.~F.}\ \bibnamefont
  {Brennan}},\ }\href {https://doi.org/10.1063/1.344408} {\bibfield  {journal}
  {\bibinfo  {journal} {J. Appl. Phys}\ }\textbf {\bibinfo {volume} {66}},\
  \bibinfo {pages} {1494} (\bibinfo {year} {1989})}\BibitemShut {NoStop}%
\bibitem [{\citenamefont {Jian}\ \emph {et~al.}(2018)\citenamefont {Jian},
  \citenamefont {Bai}, \citenamefont {Cui}, \citenamefont {Wei}, \citenamefont
  {Liu}, \citenamefont {Zhang}, \citenamefont {Sang},\ and\ \citenamefont
  {Zhang}}]{Jian2018}%
  \BibitemOpen
  \bibfield  {author} {\bibinfo {author} {\bibfnamefont {A.}~\bibnamefont
  {Jian}}, \bibinfo {author} {\bibfnamefont {G.}~\bibnamefont {Bai}}, \bibinfo
  {author} {\bibfnamefont {Y.}~\bibnamefont {Cui}}, \bibinfo {author}
  {\bibfnamefont {C.}~\bibnamefont {Wei}}, \bibinfo {author} {\bibfnamefont
  {X.}~\bibnamefont {Liu}}, \bibinfo {author} {\bibfnamefont {Q.}~\bibnamefont
  {Zhang}}, \bibinfo {author} {\bibfnamefont {S.}~\bibnamefont {Sang}},\ and\
  \bibinfo {author} {\bibfnamefont {X.}~\bibnamefont {Zhang}},\ }\href
  {https://doi.org/10.1016/j.optcom.2018.07.047} {\bibfield  {journal}
  {\bibinfo  {journal} {Optics Communications}\ }\textbf {\bibinfo {volume}
  {428}},\ \bibinfo {pages} {191} (\bibinfo {year} {2018})}\BibitemShut
  {NoStop}%
\bibitem [{\citenamefont {Berreman}(1972)}]{Berreman1972}%
  \BibitemOpen
  \bibfield  {author} {\bibinfo {author} {\bibfnamefont {D.~W.}\ \bibnamefont
  {Berreman}},\ }\href@noop {} {\bibfield  {journal} {\bibinfo  {journal}
  {Journal of the Optical Society of America}\ }\textbf {\bibinfo {volume}
  {62}},\ \bibinfo {pages} {502} (\bibinfo {year} {1972})}\BibitemShut
  {NoStop}%
\bibitem [{\citenamefont {Rogers}\ \emph {et~al.}(2011)\citenamefont {Rogers},
  \citenamefont {Kotelyanskii},\ and\ \citenamefont {Sirenko}}]{Rogers2011}%
  \BibitemOpen
  \bibfield  {author} {\bibinfo {author} {\bibfnamefont {P.}~\bibnamefont
  {Rogers}}, \bibinfo {author} {\bibfnamefont {M.}~\bibnamefont
  {Kotelyanskii}},\ and\ \bibinfo {author} {\bibfnamefont {A.}~\bibnamefont
  {Sirenko}},\ }\href@noop {} {\bibfield  {journal} {\bibinfo  {journal}
  {arXiv.org}\ } (\bibinfo {year} {2011})}\BibitemShut {NoStop}%
\bibitem [{\citenamefont {Figotin}\ and\ \citenamefont
  {Vitebsky}(2001)}]{Figotin2001}%
  \BibitemOpen
  \bibfield  {author} {\bibinfo {author} {\bibfnamefont {A.}~\bibnamefont
  {Figotin}}\ and\ \bibinfo {author} {\bibfnamefont {I.}~\bibnamefont
  {Vitebsky}},\ }\href {https://doi.org/10.1103/PhysRevE.63.066609} {\bibfield
  {journal} {\bibinfo  {journal} {Phys. Rev. E}\ }\textbf {\bibinfo {volume}
  {63}},\ \bibinfo {pages} {066609} (\bibinfo {year} {2001})}\BibitemShut
  {NoStop}%
\bibitem [{\citenamefont {Figotin}\ and\ \citenamefont
  {Vitebskiy}(2003)}]{Figotin2003}%
  \BibitemOpen
  \bibfield  {author} {\bibinfo {author} {\bibfnamefont {A.}~\bibnamefont
  {Figotin}}\ and\ \bibinfo {author} {\bibfnamefont {I.}~\bibnamefont
  {Vitebskiy}},\ }\href {https://doi.org/10.1103/PhysRevB.67.165210} {\bibfield
   {journal} {\bibinfo  {journal} {Phys. Rev. B}\ }\textbf {\bibinfo {volume}
  {67}},\ \bibinfo {pages} {165210} (\bibinfo {year} {2003})}\BibitemShut
  {NoStop}%
\bibitem [{\citenamefont {Claude}\ \emph {et~al.}(1977)\citenamefont {Claude},
  \citenamefont {Bernard},\ and\ \citenamefont {Franck}}]{Cohen-Tannoudji1977}%
  \BibitemOpen
  \bibfield  {author} {\bibinfo {author} {\bibfnamefont {C.-T.}\ \bibnamefont
  {Claude}}, \bibinfo {author} {\bibfnamefont {D.}~\bibnamefont {Bernard}},\
  and\ \bibinfo {author} {\bibfnamefont {L.}~\bibnamefont {Franck}},\
  }\href@noop {} {\emph {\bibinfo {title} {Quantum Mechanics}}}\ (\bibinfo
  {publisher} {Wiley-Interscience},\ \bibinfo {address} {New York, NY},\
  \bibinfo {year} {1977})\ pp.\ \bibinfo {pages} {69--78}\BibitemShut {NoStop}%
\bibitem [{\citenamefont {Cahay}\ and\ \citenamefont
  {Bandyopadhyay}(2017)}]{Cahay2017a}%
  \BibitemOpen
  \bibfield  {author} {\bibinfo {author} {\bibfnamefont {M.}~\bibnamefont
  {Cahay}}\ and\ \bibinfo {author} {\bibfnamefont {S.}~\bibnamefont
  {Bandyopadhyay}},\ }\href@noop {} {\emph {\bibinfo {title} {Problem Solving
  in Quantum Mechanics}}}\ (\bibinfo  {publisher} {John Wiley \& Sons},\
  \bibinfo {address} {Amsterdam, Netherlands},\ \bibinfo {year} {2017})\ p.\
  \bibinfo {pages} {368}\BibitemShut {NoStop}%
\bibitem [{\citenamefont {Zhu}\ \emph {et~al.}(1986)\citenamefont {Zhu},
  \citenamefont {Yu}, \citenamefont {Hawley},\ and\ \citenamefont {Roy}}]{zhu}%
  \BibitemOpen
  \bibfield  {author} {\bibinfo {author} {\bibfnamefont {D.}~\bibnamefont
  {Zhu}}, \bibinfo {author} {\bibfnamefont {A.~W.}\ \bibnamefont {Yu}},
  \bibinfo {author} {\bibfnamefont {D.}~\bibnamefont {Hawley}},\ and\ \bibinfo
  {author} {\bibfnamefont {R.}~\bibnamefont {Roy}},\ }\href
  {https://doi.org/10.1119/1.14514} {\bibfield  {journal} {\bibinfo  {journal}
  {American Journal of Physics}\ }\textbf {\bibinfo {volume} {54}},\ \bibinfo
  {pages} {601} (\bibinfo {year} {1986})}\BibitemShut {NoStop}%
\bibitem [{\citenamefont {Tai}\ \emph {et~al.}(1984)\citenamefont {Tai},
  \citenamefont {Kiang},\ and\ \citenamefont {Chen}}]{tai}%
  \BibitemOpen
  \bibfield  {author} {\bibinfo {author} {\bibfnamefont {G.~C.}\ \bibnamefont
  {Tai}}, \bibinfo {author} {\bibfnamefont {Y.~W.}\ \bibnamefont {Kiang}},\
  and\ \bibinfo {author} {\bibfnamefont {C.~H.}\ \bibnamefont {Chen}},\ }\href
  {https://doi.org/10.1109/TMTT.1984.1132621} {\bibfield  {journal} {\bibinfo
  {journal} {IEEE Trans. on Microw. Theory and Techn.}\ }\textbf {\bibinfo
  {volume} {32}},\ \bibinfo {pages} {111} (\bibinfo {year} {1984})}\BibitemShut
  {NoStop}%
\bibitem [{\citenamefont {Dragila}\ \emph {et~al.}(1985)\citenamefont
  {Dragila}, \citenamefont {Luther-Davies},\ and\ \citenamefont
  {Vukovic}}]{Dragila}%
  \BibitemOpen
  \bibfield  {author} {\bibinfo {author} {\bibfnamefont {B.}~\bibnamefont
  {Dragila}}, \bibinfo {author} {\bibfnamefont {B.}~\bibnamefont
  {Luther-Davies}},\ and\ \bibinfo {author} {\bibfnamefont {S.}~\bibnamefont
  {Vukovic}},\ }\href {https://doi.org/10.1103/PhysRevLett.55.1117} {\bibfield
  {journal} {\bibinfo  {journal} {Phys. Rev. Lett.}\ }\textbf {\bibinfo
  {volume} {55}},\ \bibinfo {pages} {1117} (\bibinfo {year}
  {1985})}\BibitemShut {NoStop}%
\bibitem [{\citenamefont {Liu}\ and\ \citenamefont {Behdad}(2012)}]{Liu}%
  \BibitemOpen
  \bibfield  {author} {\bibinfo {author} {\bibfnamefont {C.-H.}\ \bibnamefont
  {Liu}}\ and\ \bibinfo {author} {\bibfnamefont {N.}~\bibnamefont {Behdad}},\
  }\href {https://doi.org/10.2528/PIERB12051016} {\bibfield  {journal}
  {\bibinfo  {journal} {Progress In Electromagnetics Research B}\ }\textbf
  {\bibinfo {volume} {42}},\ \bibinfo {pages} {1} (\bibinfo {year}
  {2012})}\BibitemShut {NoStop}%
\bibitem [{\citenamefont {Henderson}\ \emph {et~al.}(1991)\citenamefont
  {Henderson}, \citenamefont {Gaylord},\ and\ \citenamefont {Glytsis}}]{hend}%
  \BibitemOpen
  \bibfield  {author} {\bibinfo {author} {\bibfnamefont {G.}~\bibnamefont
  {Henderson}}, \bibinfo {author} {\bibfnamefont {T.~K.}\ \bibnamefont
  {Gaylord}},\ and\ \bibinfo {author} {\bibfnamefont {E.}~\bibnamefont
  {Glytsis}},\ }\href {https://doi.org/10.1109/5.118988} {\bibfield  {journal}
  {\bibinfo  {journal} {Proc. IEEE}\ }\textbf {\bibinfo {volume} {79}},\
  \bibinfo {pages} {1643} (\bibinfo {year} {1991})}\BibitemShut {NoStop}%
\bibitem [{\citenamefont {Brown}(1953)}]{brown}%
  \BibitemOpen
  \bibfield  {author} {\bibinfo {author} {\bibfnamefont {J.}~\bibnamefont
  {Brown}},\ }\href {https://doi.org/10.1049/pi-4.1953.0009} {\bibfield
  {journal} {\bibinfo  {journal} {Proceedings of the IEE-Part IV: Institution
  Monographs}\ }\textbf {\bibinfo {volume} {100}},\ \bibinfo {pages} {51}
  (\bibinfo {year} {1953})}\BibitemShut {NoStop}%
\bibitem [{\citenamefont {Dung}\ \emph {et~al.}(2000)\citenamefont {Dung},
  \citenamefont {Kn\"oll},\ and\ \citenamefont {Welsch}}]{los}%
  \BibitemOpen
  \bibfield  {author} {\bibinfo {author} {\bibfnamefont {H.~T.}\ \bibnamefont
  {Dung}}, \bibinfo {author} {\bibfnamefont {L.}~\bibnamefont {Kn\"oll}},\ and\
  \bibinfo {author} {\bibfnamefont {D.-G.}\ \bibnamefont {Welsch}},\ }\href
  {https://doi.org/10.1103/PhysRevA.62.053804} {\bibfield  {journal} {\bibinfo
  {journal} {Phys. Rev. A}\ }\textbf {\bibinfo {volume} {62}},\ \bibinfo
  {pages} {053804} (\bibinfo {year} {2000})}\BibitemShut {NoStop}%
\bibitem [{\citenamefont {Nicolson}\ and\ \citenamefont {Ross}(1970)}]{nic}%
  \BibitemOpen
  \bibfield  {author} {\bibinfo {author} {\bibfnamefont {A.~M.}\ \bibnamefont
  {Nicolson}}\ and\ \bibinfo {author} {\bibfnamefont {G.~F.}\ \bibnamefont
  {Ross}},\ }\href {https://doi.org/10.1109/TIM.1970.4313932} {\bibfield
  {journal} {\bibinfo  {journal} {Trans. Instrum. Meas.}\ }\textbf {\bibinfo
  {volume} {19}},\ \bibinfo {pages} {377} (\bibinfo {year} {1970})}\BibitemShut
  {NoStop}%
\bibitem [{\citenamefont {Weir}(1974)}]{weir}%
  \BibitemOpen
  \bibfield  {author} {\bibinfo {author} {\bibfnamefont {W.~B.}\ \bibnamefont
  {Weir}},\ }\href {https://doi.org/10.1109/PROC.1974.9382} {\bibfield
  {journal} {\bibinfo  {journal} {Proc. IEEE}\ }\textbf {\bibinfo {volume}
  {62}},\ \bibinfo {pages} {33} (\bibinfo {year} {1974})}\BibitemShut {NoStop}%
\bibitem [{\citenamefont {Smith}\ \emph {et~al.}(2000)\citenamefont {Smith},
  \citenamefont {Padilla}, \citenamefont {Vier}, \citenamefont {Nemat-Nasser},\
  and\ \citenamefont {Schultz}}]{smith-ext}%
  \BibitemOpen
  \bibfield  {author} {\bibinfo {author} {\bibfnamefont {D.~R.}\ \bibnamefont
  {Smith}}, \bibinfo {author} {\bibfnamefont {W.~J.}\ \bibnamefont {Padilla}},
  \bibinfo {author} {\bibfnamefont {D.~C.}\ \bibnamefont {Vier}}, \bibinfo
  {author} {\bibfnamefont {S.~C.}\ \bibnamefont {Nemat-Nasser}},\ and\ \bibinfo
  {author} {\bibfnamefont {S.}~\bibnamefont {Schultz}},\ }\href
  {https://doi.org/10.1103/PhysRevLett.84.4184} {\bibfield  {journal} {\bibinfo
   {journal} {Phys. Rev. Lett.}\ }\textbf {\bibinfo {volume} {84}},\ \bibinfo
  {pages} {4184} (\bibinfo {year} {2000})}\BibitemShut {NoStop}%
\end{thebibliography}
\end{document}